\documentclass[aps,prl,reprint,superscriptaddress]{revtex4-2}
\usepackage{amsmath,mathrsfs,amssymb,mathtools}
\usepackage[colorlinks,linkcolor=blue,citecolor=blue,urlcolor=blue]{hyperref}

\begin{document}

\title{AE-driven Zonal Modes Produce Transport Barriers and Heat Thermal Ions by Cross-Scale Interactions}

\author{Qinghao Yan}
\email{qinghaoyan@outlook.com}
\affiliation{Center for Fusion Sciences, Southwestern Institute of Physics, Chengdu, Sichuan 610041, People's Republic of China}
\author{P. H. Diamond}
\affiliation{Department of Astronomy and Astrophysics and Department of Physics, University of California San Diego, La Jolla, California 92093, USA}
\date{\today}
\begin{abstract}
    In scenarios where a sustained energetic particle source strongly drives toroidal Alfvén eigenmodes (TAE), and phase-space transport is insufficient to saturate TAE, this novel theory of TAE-zonal mode (ZM)-turbulence -- self-regulated by cross-scale interactions (including collisionless ZF damping) -- merits consideration. Zonal modes are driven by Reynolds and Maxwell stresses, without the onset of modulational instability. TAE evolution in the presence of ZMs conserves energy and closes the system feedback loop. The saturated zonal shears can be sufficient to suppress ambient drift-ITG turbulence, achieving an enhanced core confinement regime. The necessary mechanism is identified. The saturated state is regulated by linear and turbulent zonal flow drag. This regulation leads to bursty TAE spectral oscillations, which overshoot while approaching saturation. Heating by both collisional and collisionless ZM damping deposits alpha particle energy into the thermal plasma, achieving effective alpha channeling. This theory offers a mechanism for EP-induced transport barrier formation, and predicts a novel thermal ion heating mechanism.
\end{abstract}

\maketitle

Energetic particles (EP) are known to excite instabilities such as toroidal Alfv\'{e}n eigenmodes (TAE) and energetic particle modes (EPM), potentially leading to significant EP transport and so affecting fusion plasma confinement\cite{chen_physics_2016}. However, recent experiments yielded results that contradict this expectation, showing that EP can be associated with \emph{improved thermal plasma confinement}, through the formation of internal transport barriers (ITB)\cite{di_siena_new_2021,han_sustained_2022,mazzi_enhanced_2022,citrin_overview_2023,garcia_stable_2024-1}. These findings highlight discrepancies between experimental results and previous theoretical predictions for EP physics. In addition, conventional turbulent transport models don't account for the effect of EP on thermal confinement. To address these discrepancies, mechanisms have been proposed, such as fast particle dilution\cite{hahm_fast_2023,choi_how_2023}, electromagnetic stabilization\cite{romanelli_fast_2010} and zonal mode (ZM) stabilization\cite{zhang_nonlinear_2013, spong_nonlinear_2021, mazzi_gyrokinetic_2022,di_siena_electromagnetic_2019}. ZM can address two key problems: the saturation of EP-driven instabilities via ZM\cite{mazzi_enhanced_2022,brochard_saturation_2024} to prevent strong EP transport, and the suppression of turbulence by sheared ZM\cite{diamond_zonal_2005} to form a transport barrier. The TAE/ZM system is an ideal framework for studying these issues. TAE can spontaneously generate ZM via modulational instability, which requires a threshold TAE amplitude\cite{chen_nonlinear_2012}. Conversely, the directly driven (forced-driven) process\cite{diamond_theory_1991,qiu_effects_2016,qiu_nonlinear_2017,todo_nonlinear_2010,chen_zonal_2018,brochard_saturation_2024} does not require a threshold TAE amplitude. The scenario of ZM drive by waves is a well-known paradigm for pattern and structure formation, such as atmospheric jets, Jovian bands and drift wave-zonal flow (DW-ZF) turbulence in tokamaks. Here, the new element is the presence of \emph{two} wave populations - the TAEs and DWs - which couple via the ZM. The nonlinear cross-transport in fusion plasma has been studied in Refs.\cite{qiu_gyrokinetic_2023,falessi_nonlinear_2023,spong_nonlinear_2021,biancalani_effect_2020,biancalani_gyrokinetic_2021,liu_simple_2023}, but these studies lack saturation mechanism for the directly driven process, especially the proper mechanisms for ZM collisionless damping.
\begin{figure}[b]
    \centering
    \includegraphics[width=\linewidth]{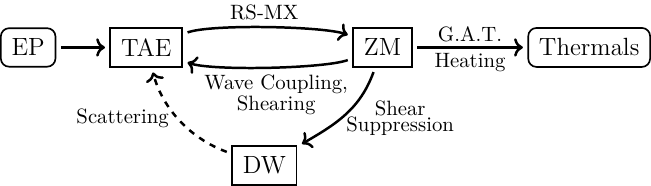}
    \caption{A feedback diagram for TAE and ZM with energy deposition from EP to thermals. ZM saturates TAE and suppresses DW turbulence. Thermal plasma damps ZM and so is heated.}\label{F:PPmodel}
\end{figure}

As we will show in this paper, with the sustained EP profile and strongly excited TAE, the interactions between TAEs and directly driven ZM lead to an interesting self-organized state of the system, which arises from the feedback loop structure as shown in Figure \ref{F:PPmodel}: ZM is directly driven by TAE via Reynolds and Maxwell stresses, and damped by collisional and collisionless drag processes. This, in turn, influences the evolution of the TAE via wave-ZM interaction. ZM damping regulates the TAE saturation level and the oscillations of TAE and ZM as they approach saturation. The geodesic acoustic transference (GAT) is a relevant collisionless damping mechanism, which requires sufficient turbulent mixing to be effective. GAT heats ions nonlinearly, ultimately functioning as an alpha channeling process. EP transport is connected to thermal turbulence by cross-scale interactions. The shearing effect from the saturated ZM can be sufficient to suppress DW turbulence, leading to a transport barrier for thermals.

To investigate the feedback interactions between TAE and ZM, we use $ \delta\phi $ and $ \delta A_{\parallel} $ for scalar potential and parallel vector potential respectively, and define $ \delta L\equiv \delta\phi-(v_{\parallel}\delta A_{\parallel}/c) $, $ \delta\psi\equiv \delta A_{\parallel}\omega /(ck_{\parallel}) $. Following the notations in Refs.\cite{qiu_effects_2016,qiu_nonlinear_2017}, let $ \delta \phi=\delta \phi_Z +\delta \phi_T $, where $ \delta\phi_Z $ is the zonal component and $ \delta\phi_T\equiv \delta\phi_0+ \delta\phi_{0*}$, with $ \delta\phi_0 $ and $ \delta\phi_{0*} $ having the opposite real frequencies. Therefore, a collision of two counter-propagating eigenmodes is considered here. Using the ballooning decomposition in $ (r,\theta,\phi) $, we have
\begin{align*}
   &\delta\phi_0=\hat{A}_0e^{i\int\hat{k}_{r,0}\textrm{d}r+i(n\phi-m_0\theta-\omega_0t)}{\textstyle\sum}_{j}e^{-ij\theta}\Phi_0(x-j)\\
    &\delta\phi_Z=\hat{A}_Ze^{i\int\hat{k}_{Z}\textrm{d}r-i\omega_Zt}{\textstyle\sum}_{m}\Phi_Z
\end{align*}
where $ \Phi_0 $ above accounts for the fine scale structure of TAE, $\hat{A}_0$ is the envelope amplitude, and $ \hat{k}_{r,0} $ stands for the radial envelope wave number. From the gyrokinetic nonlinear vorticity equation\cite{zonca_theory_2014,qiu_effects_2016,qiu_nonlinear_2017,chen_physics_2016}, one can obtain the nonlinear evolution of TAE $ \delta\phi_0 $ as equation \eqref{E:TAEevo}:
\begin{equation}\label{E:TAEevo}
    \begin{split}
        &-k_{\parallel}^2\delta\psi_0+\dfrac{\omega_0^2}{V_A^2}\delta\phi_0=\dfrac{4\pi\omega_0 e}{c^2k_{\perp,0}^2}\langle J_{k_0}\omega_d\delta H_{k_0}\rangle_{\mathcal{E}}\\
        &-i\dfrac{c}{B_0}\dfrac{k_Zk_{\theta,0}}{k_{\perp,0}^2}(k_Z^2-k_{\perp,0}^2)\dfrac{\omega_0}{V_A^2}\delta\phi_0(\delta\phi_Z-\delta\psi_Z)
    \end{split}
\end{equation}
where $ \mathcal{E}\equiv v^2/2 $, $ \mathbf{k}\delta\phi=[k_{\parallel}\hat{b}+k_{\theta}\hat{\theta}+(\hat{k}_r-inq'\partial_x\ln\Phi)\hat{r}]\delta\phi $, $V_A^{-2}\equiv 4\pi n_0 e^2 \rho_i^2/(c^2 T_i)$, $ J_{k}=J_{0}(k_{\perp}\rho_L) $ is the Bessel function, EP distribution function is $ F=F_0+\delta H $. The ZM contributions are the second term on the R.H.S. The breaking of the ideal MHD condition is expressed as:
\begin{equation}\label{E:iMHDcondition}
    \delta \phi_0-\delta \psi_0=i\dfrac{c}{B_0}\dfrac{k_{\theta}k_Z}{\omega_0}(\delta \psi_Z-\delta \phi_Z)\delta\phi_0
\end{equation}
In equation \eqref{E:TAEevo}, the integral $ \langle\cdots\rangle_{\mathcal{E}}\equiv \int\sqrt{\mathcal{E}}\text{d}\mathcal{E} $ of the fast particle distribution $ \delta H $ appears. This is obtained from the nonlinear gyrokinetic equation\cite{zonca_nonlinear_2015}:
\begin{equation}\label{E:nonlinearGK}
    \begin{split}
        &\left(-i\omega+v_{\parallel}\partial_l+i\omega_d+\Omega_Z\right)\delta H_k =\\
        &\qquad -i\dfrac{e_s}{m}Q_0F_0J_k\delta L_k-\dfrac{c}{B_0}\Lambda_k J_{k'}\delta L_{k'}\delta H_{k''}
    \end{split} 
\end{equation}
where $\Lambda_{k}\equiv \sum_{k=k'+k''}\hat{b}\cdot\mathbf{k}''\times\mathbf{k}' $. Then the ``linear'' response to the TAE mode $ \delta \phi_0 $ is written as\cite{fu_excitation_1992,biglari_resonant_1992-1}:
\begin{equation}\label{E:linearResponse}
    \delta H_0^{L}=-\dfrac{e}{m}Q_0F_0e^{i\lambda_{d0}}J_{k_0}\delta L_0 \sum_{l}\dfrac{(-1)^lJ_l(\hat{\lambda}_{d0})e^{il(\theta-\theta_0)}}{\omega_0-k_{\parallel,0}v_{\parallel}-l\omega_{t}-i\Omega_Z}
\end{equation}
where $ \omega_d=\hat{v}_{d,\mathcal{E}}(k_r\sin \theta+k_{\theta}\cos \theta) $, $ \hat{v}_{d,\mathcal{E}}=(v_{\perp}^2+2v_{\parallel}^2)/(2\Omega_i R_0) $, $ Q_0F_0\equiv (\omega_0\partial_{\mathcal{E}}-\omega_*)F_0 $,  $ \omega_*F_0=\mathbf{k}\cdot\mathbf{b}\times\nabla F_0/\Omega_i $, $ L_{n0E}^{-1}\equiv -\text{d}\ln n_{0E}/\text{d}r $, $ \Omega_i $ is the thermal ion cyclotron frequency, $ \Omega_Z\equiv ck_{\theta,0}k_ZJ_Z\delta L_Z/B_0\simeq k_{\theta,0}V_{E\times B} $ is the nonlinear phase shift from ZM, $ \lambda_{d0}=\hat{\lambda}_{d0}\sin(\theta-\theta_0) $, $ \hat{\lambda}_{d0}=k_{\perp,0}\hat{\rho}_{d,\mathcal{E}} $, $ \hat{\rho}_{d,\mathcal{E}}=qR_0\hat{v}_{d,\mathcal{E}}/v_{\parallel} $, $ \tan\theta_0=k_r/k_{\theta} $, $ J_l(\hat{\lambda}_{d0}) $ is the $ l $-th order Bessel function of $ \hat{\lambda}_{d0} $. For large parallel velocity, there is $ \omega_{t}=v_{\parallel}/qR_0 $.

Multiplying equation \eqref{E:TAEevo} by $ \delta\phi_{0*} $, utilizing equations \eqref{E:iMHDcondition} and \eqref{E:linearResponse}, and assuming a slowing down distribution function $ F_{0E}=n_{0E}\mathcal{E}^{-3/2}/\ln(\mathcal{E}_f/\mathcal{E}_c) $ with $ \mathcal{E}_c\leq\mathcal{E}\leq\mathcal{E}_f $ and $ \mathcal{E}_f=v_f^2/2 $, we have the evolution equation for $ |\delta\phi_0|^2 $ or, equivalently, a wave kinetic equation. This is (see Supplemental Material for more details):
\begin{equation}\label{E:waveEvoFull}
    \begin{split}
        &\bigg(1-\dfrac{k_{\parallel,0}^2V_A^2}{\omega_0^2}\bigg) \partial_t |\delta\phi_0|^2= \\
        &\bigg(1-\dfrac{k_{\parallel,0}V_A}{\omega_0}\bigg)\dfrac{\omega_A\pi}{2\sigma}q^3k_{\theta,0}\rho_A\beta_Ef_r\dfrac{R_0}{L_{n0E}}|\delta\phi_0|^2\\
        &-\bigg(1-\dfrac{k_Z^2}{k_{\perp,0}^2}-\dfrac{k_{\parallel,0}^2V_A^2}{\omega_0^2}\bigg)\dfrac{2c}{B_0}k_{\theta,0}k_Z|\delta\phi_0|^2\delta \phi_Z
    \end{split}
\end{equation}
Here we have only considered $ l=\pm 1 $ transit resonances, $ \omega_A\equiv V_A/(qR_0)    $, $ \rho_A\equiv V_A/\Omega_i $, $ \beta_{E}\equiv 8\pi n_{0E}m_i\mathcal{E}_f/B_0^2 $, $ \sigma\equiv \ln(\mathcal{E}_f/\mathcal{E}_c)\mathcal{E}_f/V_A^2 $, $ f_r $ is a factor for the fraction of resonant particles. We neglected zonal current (ZC) here, since ZC primarily brings additional magnetic shear to the system, which should have small effect on TAE evolution\cite{todo_nonlinear_2010,zhang_nonlinear_2013,chen_nonlinear_2012,qiu_nonlinear_2017}. Equation \eqref{E:waveEvoFull} consists of linear growth from the wave-particle resonance (R.H.S. term $\propto|\delta\phi_0|^2 $)\cite{fu_excitation_1992,biglari_resonant_1992-1}, and damping from the nonlinear interaction between TAE and ZM (R.H.S. term $ \propto|\delta\phi_0|^2 \delta\phi_Z $). Therefore, equation \eqref{E:waveEvoFull} can be cast into the form of equation \eqref{E:waveEvo}, discussed later.

The evolution equation for ZM is\cite{zonca_theory_2014,qiu_effects_2016,qiu_nonlinear_2017,chen_physics_2016}:
\begin{equation}\label{E:zonalEvo_symbolic}
    \begin{split}
        &\dfrac{e^2}{T_i}\left\langle (1-J_Z^2)F_{0,i} \right\rangle_{\mathcal{E}}\partial_t\delta \phi_Z =\\
        &\qquad-{\textstyle\sum}_s \left\langle e_sJ_{Z}i\omega_d\delta H_0 \right\rangle_{\mathcal{E},Z} +(\text{RS-MX})_Z
    \end{split}
\end{equation}
On the R.H.S., the first term is the curvature coupling term (CCT), where the bracket denotes an integral in velocity space and only the nonlinear portion of the distribution can contribute\cite{qiu_effects_2016}. The second term accounts for Reynolds and Maxwell stresses (RS-MX), which are primarily thermal contributions and can be included following Refs.\cite{chen_non-linear_2001,chen_nonlinear_2012,qiu_effects_2016,qiu_nonlinear_2017}. Considering $ k_{\perp}\rho_i\ll 1 $ and $ k_{\perp}\hat{\rho}_{d,\mathcal{E}}\ll 1 $, we have the evolution of zonal potential as (see details in Supplemental Material):
\begin{equation}\label{E:zonalEvoFull}
    \partial_t\delta\phi_Z=\dfrac{ck_{\theta,0}\hat{F}}{\hat{\chi}_{iZ}B_0}\bigg[\dfrac{n_{0E}\hat{\rho}_{d,f}^2}{n_0\rho_i^2}\hat{G}+\bigg(1-\dfrac{k_{\parallel,0}^2V_A^2}{\omega_0^2}\bigg)\bigg]|\delta\phi_0|^2
\end{equation}
where $\hat{\chi}_{iZ}\equiv \chi_{iZ}/(k_Z^2\rho_i^2)\simeq 1.6 q^2/\sqrt{\varepsilon}$ is the neoclassical polarizability of ZM\cite{rosenbluth_poloidal_1998}, $ \hat{\rho}_{d,f}\equiv qv_f/\Omega_i $. The two parts in the bracket on the R.H.S. come from CCT and RS-MX, respectively. CCT is equivalent to a turbulent stress like RS-MX. Both are proportional to the TAE amplitude $ |\delta\phi_0|^2 $ and $ \hat{F}\equiv i(k_{r,0}-k_{r,0*}) $, indicating that the collision of two TAEs and the asymmetry of mode structure are necessary for ZM generation. Both exhibit polarization effects, either from gyro-motion or EP drift. Both require the breaking of ideal MHD condition $ 1-(k_{\parallel,0}^2V_A^2/\omega_0^2)\sim \varepsilon \neq 0 $ (for CCT from $ \hat{G} $). Because $ \hat{G}\propto \varepsilon f_r|\Omega_{*E}/\omega_0|$ and $ f_r<1 $, we notice that:
\begin{equation}\label{E:CCT/RSMX}
    \dfrac{\rm CCT}{\text{RS-MX}}\sim \dfrac{f_rn_{0E}}{n_0}\dfrac{\hat{\rho}_{d,f}^2}{\rho_i^2}\left|\dfrac{\Omega_{*E}}{\omega_0}\right|\sim f_rq^2\dfrac{P_E}{P_i}\left|\dfrac{\Omega_{*E}}{\omega_0}\right| \ll 1
\end{equation}
for the strongly driven scenario $ f_{\beta}\equiv P_E/P_i \sim 1$, where $ P_E= n_{0E}T_E $, $ P_i= n_0T_i$, $ T_E\equiv m_i\mathcal{E}_f $, $ \Omega_{*E}\equiv k_{\theta} L_{n_{0E}}^{-1}cT_i/eB $. Notably, with $ \hat{\omega}_{*E}\equiv \Omega_{*E}T_E/T_i $, the ratio can be written as $ \mathcal{O}(n_{E,R}\hat{\omega}_{*E}q^2/(n_0\omega_0)) $, which is different from Ref.\cite{qiu_nonlinear_2017} by a $ \varepsilon $ factor from $ \hat{G} $. Although $ |\hat{\omega}_{*E}/\omega_0|\gg 1 $, the (resonant) particle fraction of EP is small when $ f_{\beta}\sim 1 $. Consequently, RS-MX predominantly drives the generation of ZM.
 Equation \eqref{E:zonalEvoFull} will later be cast in the form of equation \eqref{E:zonalEvo}.

Using a quasi-linear approach, we can estimate the effects of TAE on the EP distribution\cite{qiu_gyrokinetic_2019, falessi_transport_2019}. Then the evolution equation of the EP profile, accounting for external EP sources such as neutral beam heating or ion cyclotron resonance heating, is formally written as below (see Supplemental Material): 
\begin{equation}\label{E:EPevo}
    \partial_t n_{0E}=-D_{\rm res}\hat{F}L_{n_{0E}}^{-1}n_{0E}+\text{Source} + \cdots
\end{equation}
where the TAE transport coefficient is:
\begin{equation}\label{E:EPDres}
    D_{\rm res}\simeq \dfrac{\pi}{2\sigma}\dfrac{k_Z^2\hat{\rho}_{d,f}^2V_Af_r}{k_{\parallel,0}}\bigg(\dfrac{\delta B_r}{B_0}\bigg)^2
\end{equation}

Now we can simplify the evolution of TAE and ZM, neglecting CCT, and rewriting equations \eqref{E:waveEvoFull} and \eqref{E:zonalEvoFull} as equations \eqref{E:waveEvo}-\eqref{E:zonalEvo}:
\begin{align}
    \partial_t |\delta\phi_0|^2&= \gamma_1|\delta\phi_0|^2-\gamma_d|\delta\phi_0|^2\delta \phi_Z\label{E:waveEvo}\\
    \hat{\chi}_{iZ}\partial_t\delta\phi_Z&= \gamma_2|\delta\phi_0|^2-\nu\delta\phi_Z\label{E:zonalEvo}
\end{align}
where,
\begin{align*}
    \gamma_1&\equiv \dfrac{\pi \omega_A}{2\sigma}q^3k_{\theta,0}\rho_A\beta_Ef_r\dfrac{R_0}{L_{n0E}}\\
    \gamma_d&\equiv \dfrac{2c}{B_0}k_{\theta,0}k_Z,\quad
    \gamma_2\equiv\dfrac{c}{B_0}k_{\theta,0}\hat{F}\varepsilon&
\end{align*}
$ \nu=(\nu_{ii}+\nu_{G}) $ is the damping of ZM. $ \nu_{ii} $ is the ion-ion collisional damping\cite{diamond_self-regulating_1994,rosenbluth_poloidal_1998} and $ \nu_G $ stands for possible collisionless damping. Here, $ \nu $ is caused by interactions in thermal plasma, therefore, it will not affect the calculation of EP-driven TAEs. Equations \eqref{E:EPevo}-\eqref{E:zonalEvo} constitute a system describing the dynamics of EP, TAE and ZM. Note that for $ \varepsilon\hat{F}/\hat{\chi}_{iZ}=k_Z $, the RS-MX terms cancel perfectly upon addition of the evolution equation for TAE and ZM amplitude! This reflects energy conservation. This system is similar to the well-known ``Predator-Prey'' system\cite{diamond_theory_1991,diamond_self-regulating_1994,diamond_zonal_2005}, as shown in Figure \ref{F:PPmodel}. Here we are focusing on the cases where EP profile is sustained by external sources, continuum damping is negligible and wave-particle trapping is unable to saturate the TAE ($ \gamma_1/\omega_0 > 10^{-2} $, see equation (4.176) in Ref.\cite{chen_physics_2016}). In essence, \emph{this system is not near marginal}. Hence, we further simplify the system by hereafter ignoring the evolution equation \eqref{E:EPevo} for EP. This simplified system manifests a closed feedback loop, featuring a robust EP source, strongly excited TAEs, and ZM generation through RS-MX. Subsequently, the ZM is damped and feeds back to the evolution of TAE through the nonlinear wave-ZM interaction, leading to the saturation in both TAE and ZM. Assuming $ \gamma_1 $, $ \gamma_2 $, $ \gamma_d $ and $ \nu $ are constants, the saturation levels are straightforwardly obtained as below, showing that the saturation of TAE is controlled by the damping of ZM.
\begin{align}
    (\delta\phi_Z)_S&=\gamma_1/\gamma_d\label{E:ZMsat-1}\\
    (|\delta\phi_0|^2)_S&=\gamma_1\nu / (\gamma_d\gamma_2)\label{E:TAEsat-1}
\end{align}

\begin{figure}[t]
    \centering
    \includegraphics[width=0.49\linewidth]{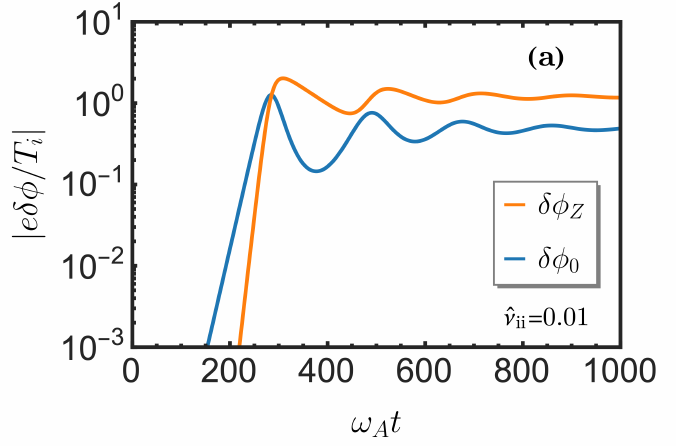}
    \includegraphics[width=0.49\linewidth]{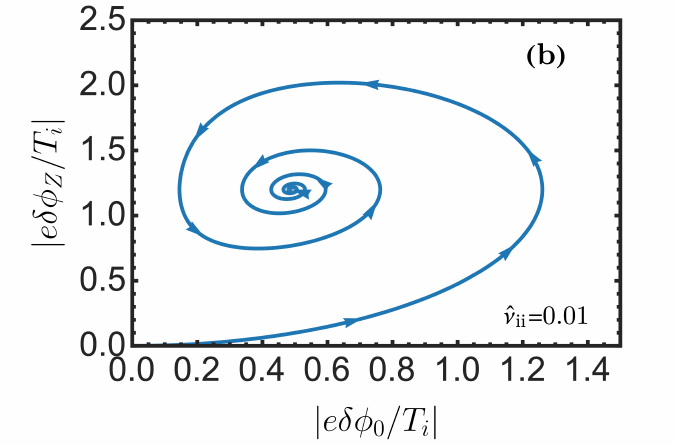}
    \includegraphics[width=0.49\linewidth]{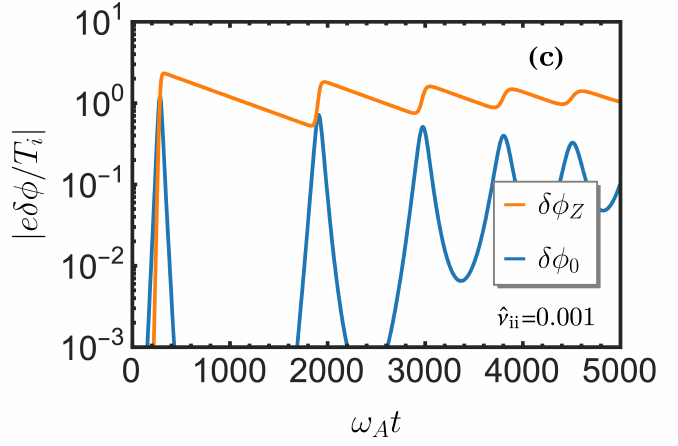}
    \includegraphics[width=0.49\linewidth]{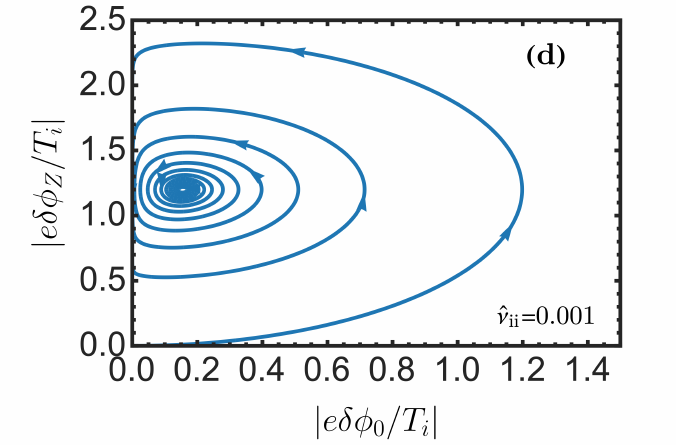}
    \includegraphics[width=0.49\linewidth]{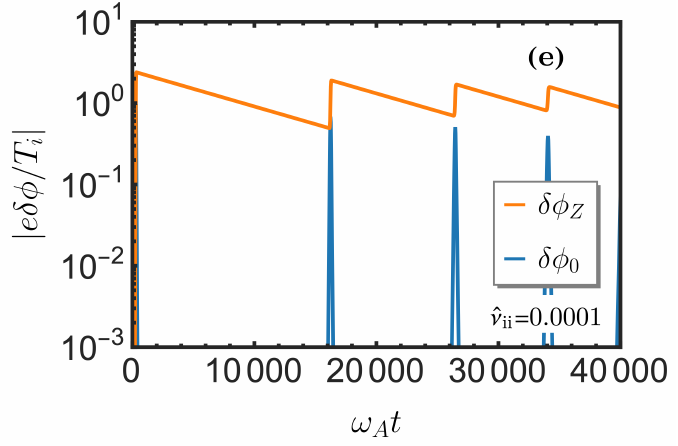}
    \includegraphics[width=0.49\linewidth]{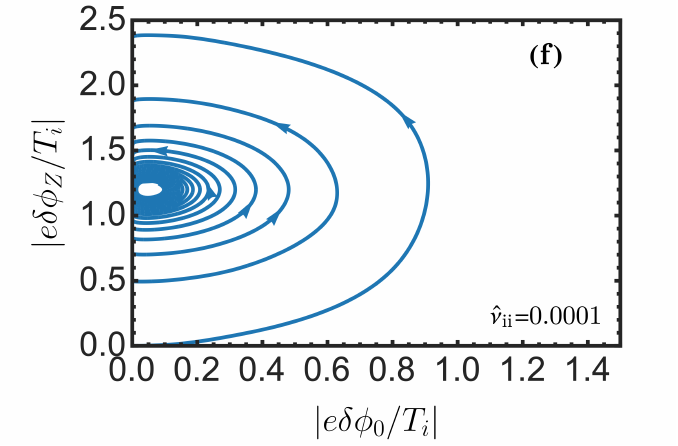}
    \caption{(a), (c), (e): TAE $ \delta\phi_0 $ and zonal potential $ \delta\phi_Z $ evolve to the fixed points in: (b), (d), (f). ZM collisional damping rate $ \hat{\nu}_{ii}\equiv\nu_{ii}/(\hat{\chi}_{iZ}\omega_A) $ is decreased from top to bottom. $ \gamma_1 $, $ \gamma_2 $ and $ \gamma_d $ are the same as in Figure \ref{F:GAT}.}
    \label{F:PPevolution}
\end{figure}

When only collisional damping is present and as $ \nu=\nu_{ii}\rightarrow 0 $, the TAE saturation level decreases, accompanied by intense oscillations in both TAE and ZM ($ f_{\rm osc}\sim \sqrt{\nu\gamma_1}/\omega_A $ when $ \nu\lesssim 4\gamma_1 $), as shown in Figure \ref{F:PPevolution}. This can appear as an intermittent (bursty) signal of TAE prior to saturation, and differs fundamentally from the quasi-periodic chirping caused by profile relaxation\cite{berk_scenarios_1992,lesur_berk-breizman_2010, spong_nonlinear_2021}. Furthermore, this property emphasizes the crucial role of collisionless ZM damping. 

Considering the significant oscillation of ZM when $ \nu_{ii}\rightarrow 0 $, which can drive thermal ion and electron oscillations in space, we introduce a simple mechanism for collisionless ZM damping, namely the geodesic acoustic transference\cite{scott_three-dimensional_1997,scott_geodesic_2003,scott_energetics_2005}. According to Refs.\cite{scott_three-dimensional_1997,scott_geodesic_2003,scott_energetics_2005}, ZF couples with a pressure sideband through the geodesic curvature coupling:
\begin{equation}\label{E:Scott1}
    \partial_t \langle \tilde{u}^{y}\rangle=\cdots-\omega_B\langle p_s \sin \vartheta\rangle
\end{equation}
where  $ \omega_B\equiv 2L_{\perp}/R $, $ L_{\perp}=(L_T,L_n) $ is the perpendicular scale of thermal profile, $ \langle \tilde{u}^{y}\rangle $ is ZF, $ \vartheta $ is the poloidal angle, $ p_s $ is the pressure sideband $ (m=1,n=0) $. The evolution of electron pressure sideband can be expressed as\cite{scott_energetics_2005}:
\begin{equation}\label{E:Scott2}
    \partial_t\langle p_{e,s} \sin \vartheta\rangle = -\tau_{\rm turb}^{-1}\langle p_{e,s} \sin \vartheta\rangle +\dfrac{\omega_B}{2} \left\langle \tilde{u}^{y}\right\rangle
\end{equation}
Here, magnetic flutter effects and ion flow sideband are neglected, defining $ \tau_{\rm turb}^{-1}\equiv D_{\rm turb}/L_{\perp}^2$. A quasilinear turbulent coefficient $ D_{\rm turb}\simeq \text{Re}\sum_{\tilde{\mathbf{k}}}i\frac{c^2}{B_0^2}|\phi_{\tilde{\mathbf{k}}}|^2/(\omega-\tilde{\mathbf{k}}_{\perp}\cdot\mathbf{v}_E) $ is assumed, with $ \tilde{\mathbf{k}}_{\perp} $ representing the wave number of the thermal turbulence and $ \mathbf{v}_{E} $ denotes $ E\times B $ advection. The energy in the pressure sideband couples to thermal particles, mostly via turbulent mixing and dissipation of parallel current\cite{scott_energetics_2005}. The resonance of the thermal particle turbulence with zonal $ E\times B $ flows gives the irreversibility which underpins the turbulent mixing. In the collisionless regime, resistivity should be weak, making turbulent mixing the dominant process.

Dedimensionalizing equations \eqref{E:waveEvo}, \eqref{E:zonalEvo} and \eqref{E:Scott2} yields equation \eqref{E:dlTAE}-\eqref{E:dlPs}, respectively. Geodesic acoustic coupling appears as the third term on the R.H.S. of equation \eqref{E:dlZF} (see Supplemental Material for detials).
\begin{align}
    \partial_\tau x &= \hat{\gamma}_1 x - \hat{\gamma}_d x y \label{E:dlTAE}\\
    \partial_\tau y &= \hat{\gamma}_2 x -\hat{\nu}_{ii} y - \hat{\gamma}_{G1} z \label{E:dlZF}\\
    \partial_\tau z &= \hat{\gamma}_{G2}y  -(\omega_A \tau_{\rm turb})^{-1} z \label{E:dlPs}
\end{align}
Here $ \tau\equiv \omega_A t $, $ x\equiv \left|e\delta\phi_0/T_i\right|^2 $, $ y\equiv e\delta\phi_Z/T_i $, $ z\equiv (1+\tau_i)\langle \hat{p}_{e,s}\sin\vartheta\rangle $, $ \tau_i\equiv T_i/T_e $, $ \hat{p}_{e,s} $ is the electron pressure sideband normalized to equilibrium pressure. Coefficients are dedimensionlized:
\begin{align*}
    &\hat{\gamma}_1 \equiv \dfrac{\gamma_1}{\omega_A}, \quad \hat{\gamma}_d\equiv \dfrac{2D_{B}k_{\theta,0}k_{Z}}{\omega_A}, \quad \hat{\gamma}_2=\dfrac{\hat{\gamma}_d}{2},\quad\hat{\nu}_{ii}\equiv \dfrac{\nu_{ii}}{\hat{\chi}_{iZ}\omega_A},\\
    &\hat{\gamma}_{G1}\equiv \dfrac{C_S^2qR_0\omega_B}{ik_ZD_B\hat{\chi}_{iZ}\omega_AL_{\perp}^2} , \quad\hat{\gamma}_{G2}\equiv (1+\tau_i)\dfrac{ i k_ZD_B}{V_A}\dfrac{\omega_B}{2}
\end{align*}
where $ D_B\equiv cT_i/(eB_0) $ and $ \varepsilon\hat{F}/\hat{\chi}_{iZ}=k_Z $ is applied. It's easy to find the geodesic acoustic oscillation from $ \hat{\gamma}_{G1}\hat{\gamma}_{G2}=\omega_{\rm GAM}^2/(\hat{\chi}_{iZ}\omega_A^2) $, where $ \omega_{\rm GAM}^2\equiv 2(1+\tau_i)C_S^2/R^2 $ is the geodesic acoustic mode (GAM) frequency, $ C_S^2\equiv T_e/M_i $. The GAT can provide extra damping of the ZM, which is obtained from the static sideband $z_{\rm static} = \hat{\gamma}_{G2} \omega_A\tau_{\rm turb} y$ and $ \hat{\gamma}_{G1} z_{\rm static}$ as:
\begin{align}\label{E:GAT-damping}
    \hat{\nu}_{G}&=\dfrac{1}{\hat{\chi}_{iZ}}\dfrac{\omega_{\rm GAM}^2}{\omega_A\tau_{\rm turb}^{-1}}\sim \mathcal{O}(10^{-2})-\mathcal{O}(10^{-1})
\end{align}
Note that turbulent mixing is necessary for collisionless ZM damping here. When compared to $\hat{\nu}_{ii}\leq\mathcal{O}(10^{-4})$, collisionless damping from GAT will be dominant.

\emph{The saturation of TAE}. Specifically, using $ k_Z^2\rho_i^2\lesssim k_{\perp}^2\rho_i^2 \sim \varepsilon^2$ and $ k_{\parallel}=1/(2qR_0) $, we obtain the saturation level of TAE from equation \eqref{E:TAEsat-1} as:
\begin{equation}\label{E:TAEsat-Estimation-1}
    \bigg(\dfrac{\delta B_r}{B_0}\bigg)_S^2 \sim \dfrac{\pi q^2}{4\sigma\hat{\chi}_{iZ}\varepsilon^2}\dfrac{\beta_Ef_r k_{\theta,0}}{L_{n0E}}\dfrac{\nu\rho_i^2}{\Omega_i}
\end{equation}
This is directly proportional to the damping of ZM, akin to the case in the DW-ZF system\cite{diamond_self-regulating_1994,diamond_zonal_2005}. Equation \eqref{E:GAT-damping} demonstrates that turbulence can regulate the saturation of TAE through the ZM damping. In addition, if EP transport is governed by TAE, since $ \hat{\nu}_G\propto \tau_{\rm turb} $, then \emph{as turbulence increases (indicated by the decrease in $ \tau_{\rm turb}$), the EP transport coefficient will decrease}. However, as we will discuss latter, the shearing feedback of ZM on turbulence will modify this straightforward conclusion. 

\begin{figure}[t]
    \centering
    \includegraphics[width=\linewidth]{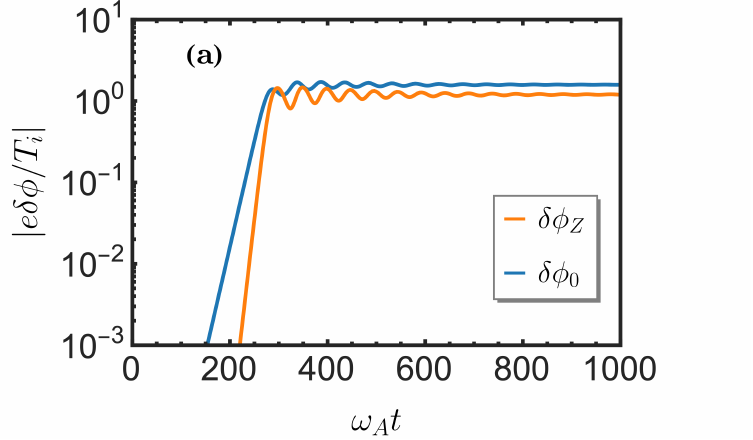}
    \includegraphics[width=\linewidth]{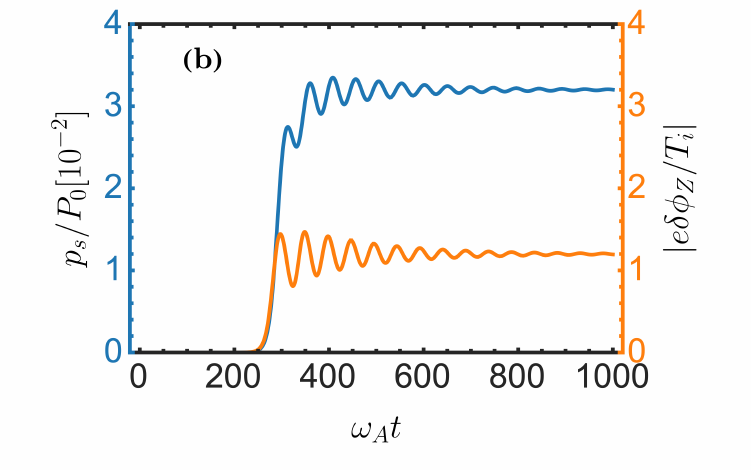}
    \caption{The evolution of TAE-ZM system with the geodesic acoustic transference as collisionless damping. The collisional damping rate is $ \hat{\nu}_{ii}=10^{-5} $, the effective collisionless damping rate is $ \hat{\nu}_{G}=10^{-1} $. Oscillations with frequency of $ \omega_{\rm GAM} $ are observed when $ \tau_{\rm Z}\sim\tau_{\rm turb} $.}
    \label{F:GAT}
\end{figure}

Introducing $ \tau_{\rm Z}^{-1}\equiv D_Bk_{\theta,0}k_Z $, we notice that when $ 3\hat{\gamma}_d\hat{\gamma}_2 \sim (\omega_A\tau_{\rm turb})^{-2}$ or $ \tau_{\rm Z}\sim\tau_{\rm turb} $, the system will oscillate at the frequency of $ \omega_{\rm GAM} $ rather than $ f_{\rm osc} $, which is similar to the results in Ref.\cite{todo_nonlinear_2010}. This condition indicates that the energy drained by ZM from the TAE equals to the energy dissipated by the pressure sideband. As a result, the energy transfer will be controlled by GAT, leading to oscillations at $ \omega_{\rm GAM} $. To generate such oscillations, we take the Bohm coefficient $ D_{\rm turb}=D_B/16 $ and $ (\rho_i/L_{\perp})^2\sim 0.02 $ in the subsequent analyses, ensuring $ \tau_{\rm Z}\sim\tau_{\rm turb} $ and $ 1/(\omega_A\tau_{\rm turb})\ll 1 $. For typical parameters in tokamaks like DIII-D with $ q = 1.5 $, $ \sigma\sim 2.5 $, $ \beta_E=1\% $, $ f_r=0.25 $, $ k_{\theta,0}=nq/a\sim 10$, $ L_{n0E}=0.5 a $, we have $ \hat{\gamma}_1\sim 0.12 $, $ \hat{\gamma}_d\sim 0.1 $, $ \hat{\gamma}_2\sim 0.05 $, $ \hat{\nu}_G\sim 0.1 $, $ \hat{\gamma}_{G1}\sim -3.9i $, $ \hat{\gamma}_{G2}\sim 0.0012i $, $ 1/(\omega_A \tau_{\text{turb}})\sim 0.045$ (see Supplementary materials for more details), then the evolution of the system is shown in Figure \ref{F:GAT}.  The saturated TAE amplitude is $ (\delta B_r/B_0)^2_S \sim 10^{-7}$. Comparing this to the threshold for spontaneous generation $ \rho_i^2/(4\varepsilon (qR_0)^2) $ \cite{chen_nonlinear_2012}, yields a ratio of $ \rm{TAE_{sat/th}}\sim \mathcal{O}(10^{-1})-\mathcal{O}(1) $. This suggests that spontaneous generation is possible after the directly driven process saturates. A similar process is found for ITG turbulence simulation\cite{wang_zonal_2017,wang_nonlinear_2022,chen_beat-driven_2024}, where eigenmode self-interaction\cite{c_j_how_2020} dominates ZF generation, and modulation of sidebands\cite{diamond_secondary_2001} becomes important only near nonlinear saturation. The estimated EP transport coefficient using equation \eqref{E:EPDres} is around $ D_{\rm res}\sim\mathcal{O}(10) \rm{m^2/s} $, which exceeds the neoclassical value but is limited, and so does not cause severe EP transport. This also facilitates the maintenance of a sustained EP profile through modulated heating, to align with the assumption of a fixed linear TAE growth rate.

\emph{The saturation of ZM}. The saturation level of zonal potential can be estimated from equation \eqref{E:ZMsat-1} as:
\begin{equation}\label{E:ZFsat-Estimation-1}
    (\delta\phi_Z)_S\sim \dfrac{\pi}{2\sigma}\dfrac{T_i}{e}\dfrac{f_{\beta}f_rq^2}{k_ZL_{n0E}}
\end{equation}
Then the ZF shear $ \omega_{E\times B}\equiv  ck_Z^2\delta\phi_Z/B_0 $ is:
\begin{align}
    \omega_{E\times B} =& \dfrac{\pi}{2\sigma} q^2D_Bf_{\beta}f_r\dfrac{k_Z}{L_{n_{0E}}}\sim \dfrac{\pi}{2\sigma} q^2 \varepsilon f_{\beta}f_r\dfrac{C_S}{L_{n_{0E}}}\label{E:ZFsat-Estimation-2}
\end{align}
The radial scale of ZM $ k_Z $ is crucial in determining the saturation levels. Factors such as the fraction of EP pressure $ f_{\beta} $, the fraction of resonant particles $ f_r $ and the EP gradient $ L_{n_{0E}}^{-1} $ also influence the strength of ZF\cite{mazzi_gyrokinetic_2022}. For the typical parameters mentioned previously, the ZF shear reaches around $ 0.3 C_S/a $ in regions where TAEs are excited, a magnitude comparable to the ITG linear growth rate with dilution effect $ \gamma_{\rm{ITG}}\sim 0.1 C_S/a $ \cite{lin_self-sustaining_2023,kim_turbulence_2023}. Therefore, the suppression of ITG turbulent transport can be anticipated. The threshold condition for shear suppression, $ \omega_{E\times B}>\gamma_{\rm ITG} $, translates into a criterion for the TAE linear growth rate $ \gamma_{\rm TAE}=\gamma_1/2>\gamma_{\rm ITG}k_{\theta,0}/k_Z $. This criterion further implies a threshold for the strength of the EP source (e.g., from NBI heating). There may appear to be a conflict concerning strong ZF shearing and necessary turbulence for ZM damping. However, since ZM damping occurs via a pressure sideband, in a strongly (weakly) turbulent state, the pressure sideband weakens (strengthens), resulting in strong (weak) ZM and shearing, and consequently facilitating reduction (enhancement) of turbulence. Moreover, when $ \tau_{\rm Z}\sim\tau_{\rm turb} $, the GAM plays a role in dissipating EP energy in this system. In general, zonal shears do affect DW and consequently regulate TAE, a process not explored in this paper. This interaction could result in a dynamical $ \hat{\nu}_G $ and thereby influence EP transport. 

The $ E\times B $ velocity caused by ZF can be estimated as:
\begin{equation}\label{E:ZFsat-Estimation-3}
    V_{E\times B}\sim \dfrac{\pi}{2\sigma}q^2f_{\beta}f_r\dfrac{C_S \rho_i}{L_{n_{0E}}}
\end{equation}
The heating power transferred from EP to thermals, estimated from equation \eqref{E:zonalEvo} as $ \nu_GV_{E\times B}^2 $ in steady state, can be distributed to both ions and electrons ($ \tau_i\sim 1 $) or predominately into electrons ($ \tau_i\ll 1 $). The ZF phase shift ($ \Omega_Z $) in equation \eqref{E:linearResponse} is smaller than the transit frequency ($ \omega_{t} $) by roughly two orders of magnitude. Thus, $ \Omega_Z $ affects near marginal transit resonance or trapped particle dynamics\cite{zonca_nonlinear_2015,brochard_saturation_2024}. Modulation of wave-particle resonance by ZF could lead mesoscopic structure such as the $ E\times B $ staircase\cite{dif-pradalier_finding_2015,yan_staircase_2022}. Additionally, a thermal ion ITB produced by ZF shearing will likely drive substantial bulk ion intrinsic rotation\cite{kosuga_efficiency_2010,ida_spontaneous_2010,fiore_production_2012,diamond_overview_2013}. 

This work introduces a novel Predator-Prey type theory for the collective dynamics of EP, TAE, turbulence and ZM, which highlights the crucial role of collisionless ZM damping and predicts a novel alpha channeling mechanism. This work presents a theoretical explanation for ITB formation driven by EPs through directly driven ZM, and a new explanation of bursty TAE phenomena. These results may contribute to the development of a novel operational scenario in future fusion devices like ITER. Other collisionless ZM damping mechanisms, such as the resonant diffusion of zonal vorticity itself\cite{li_another_2018}, are also noteworthy. The effective Richardson number is $ {\rm{Ri}}\sim (\gamma_1/\omega_{E\times B})^2 \sim (k_{\theta,0}/k_Z)^2$ and $ 0.25< \rm{Ri} \lesssim 0.5$, suggests that the K-H instability is unlikely to damp ZM for the parameters considered here\cite{Miles_1961,Howard_1961}. 

We assumed a sustained EP drive to eliminate EP profile relaxation, which is a reasonable strategy for isolating the key new physics -- the saturation of AE via collisionless ZM damping. When the system is near-marginal, the interplay between the ZF shift, wave-particle trapping\cite{fu_nonlinear_1995,todo_magnetohydrodynamic_1995,briguglio_hybrid_1998} and stochastic scattering\cite{lang_nonlinear_2011,lilley_effect_2010,duarte_theory_2017,chen_scattering_2022} is also interesting. This model is relevant to scenarios without significant phase-space transport, characterized by the absence of complex mode chirping and overlap in the spectrum. Without ZM, our model would lead to an unbounded TAE growth. The most direct correction is to consider the growth rate feedback from profile flattening or by phase-space relaxation in future. The effective EP phase space gradient reconstruction rate $ \nu_{\rm eff}\approx \nu_{\perp} \omega_A/\omega_b $ \cite{berk_scenarios_1992} with wave trapping frequency $ \omega_b\propto \sqrt{\phi_{\rm AE}} $, increases as AE saturation amplitude decreases in the presence of ZM \cite{todo_nonlinear_2010,chen_zonal_2018,biancalani_effect_2020,biancalani_gyrokinetic_2021}. In addition, turbulence may also have effects on decorrelating EP phase-space clumps and holes\cite{liu_cross-scale_2024}. Our model can serve as a basis for understanding these issues in the future. Finally, this work explores novel cross-scale interactions in fusion plasmas, introducing a compelling mechanism for transferring EP energy to thermal ions via ZM, while facilitating the formation of a TAE-driven ITB to enhance thermal confinement.

\acknowledgments
We thank Ruirui Ma, Michael Fitzgerald, T. S. Hahm, Gyungjin Choi, Xiaodi Du and Wei Chen for useful discussions. This work was supported by the National Key R\&D Program of China under Grant No. 2022YFE03100004 and National Science Foundation of China Grant No.12175053, U.S. DOE, Office of Science, under Award Number DE-FG02-04ER54738. The authors would like to thank the Isaac Newton Institute for Mathematical Sciences, Cambridge, for support and hospitality during the Program ``Anti-diffusive Dynamics: from Sub-Cellular to Astrophysical Scales'', where part of the work on this paper was undertaken. This work was supported by EPSRC grant EP/R014604/1.

\bibliographystyle{unsrt}
\bibliography{refs}

\end{document}


\title{Supplemental Material of the Letter\\``AE-driven Zonal Modes Produce Transport Barriers and Heat Thermal Ions by Cross-Scale Interactions''}

\author{Qinghao Yan}
\email{qinghaoyan@outlook.com}
\affiliation{Center for Fusion Sciences, Southwestern Institute of Physics, Chengdu, Sichuan 610041, People's Republic of China}
\author{P. H. Diamond}
\affiliation{Department of Astronomy and Astrophysics and Department of Physics, University of California San Diego, La Jolla, California 92093, USA}

\maketitle
\section{TAE evolution \label{A:TAEevo}}
The gyrokinetic nonlinear vorticity equation is\cite{zonca_theory_2014,qiu_effects_2016,chen_physics_2016}:
\begin{equation}\label{E:gkVorticity}
    \begin{split}
        &-i\dfrac{c^2B_0}{4\pi \omega_k}\partial_l\bigg(\dfrac{k_{\perp}^2}{B}\bigg)\partial_l\delta\psi_k+\dfrac{e^2}{T_i}\left\langle (1-J_k^2)F_{0,i} \right\rangle_{\mathcal{E}}\partial_t\delta\phi_k\\
        &=-\sum_s\left\langle e_sJ_k i \omega_d\delta H\right\rangle_{\mathcal{E}}-\dfrac{c\Lambda_k}{B_0}\bigg[\dfrac{c^2k^{''2}_{\perp}\partial_l\delta\psi_{k'}\partial_l\delta\psi_{k''}}{4\pi \omega_{k'}\omega_{k''}}\\
        &\qquad +\left\langle e(J_{k}J_{k'}-J_{k''})\delta L_{k'}\delta H_{k''} \right\rangle\bigg]
    \end{split}
\end{equation}
From equation \eqref{E:gkVorticity}, one can obtain the TAE evolution as represented by equation (1) of the Letter. Applying equation (2) of the Letter, the evolution of TAE amplitude is:
\begin{equation}
    \begin{split}
        &\bigg(1-\dfrac{k_{\parallel,0}^2V_A^2}{\omega_0^2}\bigg) \partial_t |\delta\phi_0|^2=-2\dfrac{4\pi e V_A^2}{c^2k_{\perp,0}^2}\left\langle J_{k_0} i \omega_d\delta H\right\rangle_{\mathcal{E}}\delta\phi_{0*}\\
        &-\bigg(1-\dfrac{k_Z^2}{k_{\perp,0}^2}-\dfrac{k_{\parallel,0}^2V_A^2}{\omega_0^2}\bigg)\dfrac{2c}{B_0}k_{\theta,0}k_Z|\delta\phi_0|^2\delta \phi_Z
    \end{split}
\end{equation}
Using the linear response equation (4) of the Letter, and notice the push-forward operator should result in the factor $ e^{-i\hat{\lambda}_{d0}\sin (\theta-\theta_0)}=\sum_l (-1)^{l}J_l(\hat{\lambda}_{d0})e^{il(\theta-\theta_{0})} $, the particle resonance drive for TAE is:
\begin{align*}
    &\left\langle J_{k_0}i\omega_d\delta H_k^{L} \right\rangle_{\mathcal{E}} &\\
    &\quad= -\bigg\langle\dfrac{e}{m}J_{k_0}^2\delta L_0 Q_0F_{0E} i \hat{v}_{d,\mathcal{E}}k_{\perp,0}\sum_l \dfrac{(-1)^lJ_l(\hat{\lambda}_{d0})}{\omega_0-k_{\parallel}v_{\parallel}-l\omega_{\rm{t}}}\\
    &\qquad\times\dfrac{1}{2\pi}\int \text{d}\theta \cos(\theta-\theta_0)e^{il(\theta-\theta_0)}e^{i\hat{\lambda}_{d0}\sin(\theta-\theta_0)}\bigg\rangle_{\mathcal{E}}\\
    &\quad \simeq \bigg\langle\dfrac{e}{m}\delta L_0 Q_0F_{0E} \hat{v}_{d,\mathcal{E}}\hat{\rho}_{d,\mathcal{E}}\dfrac{k_{\perp,0}^2}{4}\sum_{l=\pm 1} \dfrac{i\text{sign}(l)}{\omega_0-k_{\parallel}v_{\parallel}-l\omega_{\rm{t}}}\bigg\rangle_{\mathcal{E}}\\
\end{align*}
where we have used
\begin{align*}
    &\dfrac{1}{2\pi}\int \text{d}\theta \cos(\theta-\theta_0)e^{il(\theta-\theta_0)}e^{i\hat{\lambda}_{d0}\sin(\theta-\theta_0)}\\
    &=\dfrac{1}{4\pi}\int \text{d}\theta \left[e^{i(l+1)(\theta-\theta_0)}+e^{i(l-1)(\theta-\theta_0)}\right]e^{i\hat{\lambda}_{d0}\sin(\theta-\theta_0)}\\
    &=\dfrac{1}{2}\left[J_{-l-1}(\hat{\lambda}_{d0})+J_{-l+1}(\hat{\lambda}_{d0})\right]
\end{align*}
and only $ l=\pm 1 $ transit resonance are considered. Notice $ J_{-l}(a)=(-1)^{l}J_l(a) $, $ k_{\perp}\rho_L\ll 1 $ and $ k_{\perp}\rho_{d,\mathcal{E}}\ll 1 $, $ J_{k_0}=J_{0}(k_{\perp,0}\rho_L)\simeq 1$, $ J_1(\hat{\lambda}_{d0})\simeq \frac{\hat{\lambda}_{d0}}{2}$, $ J_2(\hat{\lambda}_{d0})\simeq \frac{\hat{\lambda}_{d0}^2}{8}$, $ \hat{\lambda}_{d0}=k_{\perp,0}\hat{\rho}_{d,\mathcal{E}} $. 

For large parallel velocity, we have $ v_{\parallel}\simeq \sqrt{\mathcal{E}} $ and 
\begin{equation*}
    \delta L_{0}=\delta\phi_0-\dfrac{v_{\parallel}}{c}\delta A_{\parallel,0}\simeq \delta\phi_0\left(1-\dfrac{k_{\parallel,0}v_{\parallel}}{\omega_0}\right)
\end{equation*}
Then,
\begin{align*}
    &\left\langle J_{k_0}i\omega_d\delta H_k^{L} \right\rangle_{\mathcal{E}} =\delta\phi_0\dfrac{k_{\perp,0}^2}{4}\dfrac{e}{m}\dfrac{q n_{0E}}{R_0\Omega_i^2}\bigg\langle\!\left(\!1-\dfrac{\sqrt{\mathcal{E}}k_{\parallel,0}}{\omega_0}\!\right)\times\\
    &\!\left(\!\omega_0\partial_{\mathcal{E}}+\dfrac{k_{\theta}}{\Omega_iL_{n_{0E}}}\!\right)\!\dfrac{F_{0E}}{n_{0E}} \sum_{l=\pm 1} \dfrac{i\mathcal{E}^{3/2}\text{sign}(l)}{\omega_0-(k_{\parallel}+\frac{l}{qR_0})\sqrt{\mathcal{E}}}\bigg\rangle_{\mathcal{E}}
\end{align*}
Assuming the slowing down distribution $ F_{0E}=n_{0E}\mathcal{E}^{-3/2}/\ln(\mathcal{E}_f/\mathcal{E}_c) $, then the energy integral will be:
\begin{align*}
    &\int\!\left(\!1-\dfrac{\sqrt{\mathcal{E}}k_{\parallel,0}}{\omega_0}\!\right)\!\left(\!\omega_0\partial_{\mathcal{E}}+\dfrac{k_{\theta}}{\Omega_iL_{n_{0E}}}\!\right)\!\dfrac{1}{\ln(\mathcal{E}_f/\mathcal{E}_c)}\\
    &\qquad\times\sum_{l=\pm 1} \dfrac{i\mathcal{E}^{1/2}\text{d}\mathcal{E}\text{sign}(l)}{\omega_0-(k_{\parallel}+\frac{l}{qR_0})\sqrt{\mathcal{E}}}\\
    &\simeq \left(\!1-\dfrac{k_{\parallel,0}V_A}{\omega_0}\!\right)\dfrac{1}{\ln(\mathcal{E}_f/\mathcal{E}_c)}\dfrac{k_{\theta}}{\Omega_iL_{n_{0E}}}\\
    &\quad\times\sum_{l=\pm 1}i \text{sign}(l)\bigg[\dfrac{2 \omega_0^2}{(k_{\parallel}+\frac{l}{qR_0})^3}\ln \dfrac{1-\frac{\sqrt{\mathcal{E}_c}(k_{\parallel}+\frac{l}{qR_0})}{\omega_0}}{1-\frac{\sqrt{\mathcal{E}_f}(k_{\parallel}+\frac{l}{qR_0})}{\omega_0}}\\
    &\qquad +\dfrac{2\omega_0^2(\sqrt{\mathcal{E}_c}-\sqrt{\mathcal{E}_f})}{(k_{\parallel}+\frac{l}{qR_0})^2}+\dfrac{\omega_0(\mathcal{E}_c-\mathcal{E}_f)}{k_{\parallel}+\frac{l}{qR_0}}\bigg] 
\end{align*}
Assuming there is $ \mathcal{E}_f\gtrsim V_A \gg  \mathcal{E}_c $ and $ \gamma_L\ll \omega_{0R} $, then the real part is given by:
\begin{align*}
    &\!\left(\!1-\dfrac{k_{\parallel,0}V_A}{\omega_0}\!\right)\dfrac{1}{\ln(\mathcal{E}_f/\mathcal{E}_c)}\dfrac{k_{\theta,0}}{\Omega_iL_{n_{0E}}}\\
    &\quad\times\sum_{l=\pm 1}i\dfrac{-2 \omega_0^2\text{sign}(l)}{(k_{\parallel}+\frac{l}{qR_0})^3} i \Im \ln \bigg[1-\frac{\sqrt{\mathcal{E}_f}(k_{\parallel}+\frac{l}{qR_0})}{\omega_0}\bigg]\\
    &=-\!\left(\!1-\dfrac{k_{\parallel,0}V_A}{\omega_0}\!\right)\dfrac{1}{\ln(\mathcal{E}_f/\mathcal{E}_c)}\dfrac{k_{\theta,0}}{\Omega_iL_{n_{0E}}}\\
    &\quad\times\sum_{l=\pm 1}\dfrac{2 \omega_0^2\text{sign}(l)}{(k_{\parallel}+\frac{l}{qR_0})^3} i i (-\pi) \text{Heav}(\frac{\sqrt{\mathcal{E}_f}(k_{\parallel}+\frac{l}{qR_0})}{\omega_0}-1)\\
    &=-\!\left(\!1-\dfrac{k_{\parallel,0}V_A}{\omega_0}\!\right)\dfrac{1}{\ln(\mathcal{E}_f/\mathcal{E}_c)}\dfrac{2\pi k_{\theta,0}V_A^3}{|\omega_0|\Omega_iL_{n_{0E}}}\\
    &\quad\times\left[\text{Heav}(\frac{\sqrt{\mathcal{E}_f}}{V_A}-1)+\dfrac{1}{9}\text{Heav}(\frac{3\sqrt{\mathcal{E}_f}}{V_A}-1)\right]
\end{align*}
where $ \text{Heav}(x) $ is the Heaviside Theta function, which is 0 for $ x<0 $ and 1 for $ x\geq 0 $. Finally, there is:
\begin{align*}
    &\dfrac{4\pi e V_A^2}{c^2k_{\perp,0}^2}\delta\phi_{0*} \times \delta\phi_0\dfrac{k_{\perp,0}^2}{4}\dfrac{e}{m}\dfrac{q n_{0E}}{R_0\Omega_i^2}\times (-)\times\!\left(\!1-\dfrac{k_{\parallel,0}V_A}{\omega_0}\!\right)\times\\
    &\dfrac{1}{\ln(\mathcal{E}_f/\mathcal{E}_c)}\dfrac{2\pi k_{\theta}V_A^3}{|\omega_0|\Omega_iL_{n_{0E}}}\\
    &=-\!\left(\!1-\dfrac{k_{\parallel,0}V_A}{\omega_0}\!\right)\!|\delta\phi_0|^2\dfrac{\pi \omega_A q^3 k_{\theta,0}\rho_A \beta_E}{4\ln(\mathcal{E}_f/\mathcal{E}_c)\mathcal{E}_f/V_A^2}\dfrac{R_0}{L_{n0E}}
\end{align*}
where $ V_A^{-2}\equiv 4\pi n_0e^2\rho_i^2/(c^2T_i)=4\pi \rho_m/B_0^2 $, $ \omega_A\equiv V_A/qR_0 $, $ \beta_E\equiv 8\pi n_{0E}m_i\mathcal{E}_f/B_0^2 $, $ \rho_A \equiv V_A/\Omega_i $. Putting it back into the TAE amplitude evolution gives equation (5) of the Letter as:
\begin{equation}\label{E:app:waveEvoFull}
    \begin{split}
        &\bigg(1-\dfrac{k_{\parallel,0}^2V_A^2}{\omega_0^2}\bigg) \partial_t |\delta\phi_0|^2= \\
        &\bigg(1-\dfrac{k_{\parallel,0}V_A}{\omega_0}\bigg)\dfrac{\omega_A\pi}{2 \sigma}q^3k_{\theta,0}\rho_A\beta_E f_r\dfrac{R_0}{L_{n0E}}|\delta\phi_0|^2\\
        &-\bigg(1-\dfrac{k_Z^2}{k_{\perp,0}^2}-\dfrac{k_{\parallel,0}^2V_A^2}{\omega_0^2}\bigg)\dfrac{2c}{B_0}k_{\theta,0}k_Z|\delta\phi_0|^2\delta \phi_Z
    \end{split}
\end{equation}
where $ \sigma\equiv \ln(\mathcal{E}_f/\mathcal{E}_c)\mathcal{E}_f/V_A^2 $. A factor related to the fraction of resonant particles $ f_r $ is added. Assuming $ (1-\frac{k_{\parallel}^2V_A^2}{\omega_0^2}) \sim \varepsilon $, $ (k_Z/k_{\perp,0})^2\ll 1 $, then we can have:
\begin{equation}
    \begin{split}
        \partial_t |\delta\phi_0|^2=& \gamma_1\delta\phi_0|^2-\gamma_d\delta\phi_0|^2\delta \phi_Z
    \end{split}
\end{equation}
where
\begin{align}
    \gamma_1&\equiv \dfrac{\pi \omega_A}{2\sigma}q^3k_{\theta,0}\rho_A\beta_Ef_r\dfrac{R_0}{L_{n0E}}\\
    \gamma_d&\equiv \dfrac{2c}{B_0}k_{\theta,0}k_Z
\end{align}
Then the saturation level of ZM is:
\begin{align}
    (\delta\phi_Z)_S=\dfrac{\gamma_1}{\gamma_d}&\simeq \dfrac{\pi}{2\sigma}\dfrac{T_i}{e}\dfrac{q^2f_{\beta}f_r}{k_ZL_{n0E}}
\end{align}
which is equation (22) of the Letter. Then the $ E\times B $ shear is:
\begin{align}
    \omega_{E\times B}&\equiv \dfrac{ck_Z^2}{B_0}\delta\phi_Z = \dfrac{\pi}{2\sigma}\dfrac{cT_i}{eB_0}\dfrac{q^2f_{\beta}f_rk_Z}{L_{n0E}}\notag\\
    &\sim \dfrac{\pi}{2\sigma}\dfrac{q^2f_{\beta}f_r}{L_{n0E}}C_S\rho_i \dfrac{\varepsilon}{\rho_i}=\dfrac{\pi}{2\sigma}q^2f_{\beta}f_r \varepsilon\dfrac{C_S}{L_{n0E}}
\end{align}
which is equation (23) of the Letter, where $ k_Z^2\rho_i^2<k_{\perp}^2\rho_i^2\sim \varepsilon^2 $ is used. The threshold condition $ \omega_{E\times B}>0.1C_S/a $ gives $ f_{\beta} f_r>0.1 (2\sigma L_{n0E})/(\pi \varepsilon q^2 a)\sim 0.1$, for parameters used in the Letter. The $ E\times B $ velocity is
\begin{align}
    V_{E\times B}=\dfrac{ck_Z}{B_0}\delta\phi_Z &=\dfrac{\pi}{2\sigma}q^2f_{\beta}f_r \dfrac{C_S\rho_i}{L_{n0E}}
\end{align}

\section{Zonal mode evolution \label{A:ZMevo}}
From equation \eqref{E:gkVorticity}, one can obtain the zonal mode evolution as equation (6) of the Letter. The contribution of CCT to ZM evolution is obtained from the nonlinear response\cite{qiu_effects_2016}: 
\begin{equation}
    \left(i\omega_d\delta H_0^{NL}\right)_{Z}\sim -\dfrac{c}{B_0}\Lambda_Z J_{k'}\delta L_{k'}\delta H_{k''}^{L}
\end{equation}
where $ \Lambda_Z\equiv \sum_{k_Z=k'+k''}\hat{b}\cdot\mathbf{k}''\times\mathbf{k}' $. The it gives:
\begin{gather*}
    \left\langle e_i J_Z i\omega_d\delta H_0^{NL} \right\rangle_{\mathcal{E},Z} = \dfrac{1}{2\pi}\dfrac{ce}{B_0}\times\\
    \left\langle J_Z\int d\theta \hat{H}_0 \left(J_{k_0}\delta L_{0}\delta H_{0*}^{L} - J_{k_{0*}}\delta L_{0*}\delta H_{0}^{L}\right)\right\rangle_{\mathcal{E}}
\end{gather*}
Then
\begin{align*}
    J_{k_0}\delta L_0\delta H_{0*}^{L}&=-\dfrac{e}{m}J_{k_0}J_{k_{0*}}\delta\phi_0\delta\phi_{0*}\times\\
    &\quad\left(1-\dfrac{k_{\parallel}v_{\parallel}}{\omega}\right)_{0}\left(1-\dfrac{k_{\parallel}v_{\parallel}}{\omega}\right)_{0*}Q_{0*}F_0\times\\
    &\quad\sum_{l*}\dfrac{(-1)^{l_{*}}J_{l*}(\hat{\lambda}_{d0*})e^{il_{*}(\theta-\theta_{0*})}e^{i\lambda_{d0*}}}{\omega_{0*}-k_{\parallel,0*}v_{\parallel}-l_*\omega_{\rm{t}}-i\Omega_{Z*}}\\
    J_{k_{0*}}\delta L_{0*}\delta H_{0}^{L}&=-\dfrac{e}{m}J_{k_0}J_{k_{0*}}\delta\phi_0\delta\phi_{0*}\times\\
    &\quad\left(1-\dfrac{k_{\parallel}v_{\parallel}}{\omega}\right)_{0}\left(1-\dfrac{k_{\parallel}v_{\parallel}}{\omega}\right)_{0*}Q_{0}F_0\times\\
    &\quad\sum_{l}\dfrac{(-1)^{l}J_{l}(\hat{\lambda}_{d0})e^{il(\theta-\theta_{0})}e^{i\lambda_{d0}}}{\omega_{0}-k_{\parallel,0}v_{\parallel}-l\omega_{\rm{t}}-i\Omega_{Z}}
\end{align*}
Notice $ Q_0F_0=(\omega_0\partial_E-\omega_{*})F_0=-Q_{0*}F_0 $ and $ \hat{H}_0\equiv k_{\theta,0}(k_{r,0}+k_{r,0*}) =-\hat{H}_{0*} $, $ \theta_{0*}\equiv \tanh^{-1}(k_{r,0*}/k_{\theta,0*}) $. The sign of CCT should not change when exchange $ \omega_{0} $ and $ \omega_{0*} $. Therefore $ l_*=-l $. Using the relations $ e^{ia\sin\theta}=\sum_{l}J_l(a)e^{il\theta} $, $ J_{-l}(a)=(-1)^{l}J_l(a) $, $ \lambda_{d0}= \hat{\lambda}_{d0}\sin(\theta-\theta_0)$, we have
\begin{align*}
    &\int\text{d}\theta (-1)^{l}J_{l}(\hat{\lambda}_{d0}) e^{il(\theta-\theta_{0})}e^{i\lambda_{d0}}\\
    & = \sum_{l'}\int_{-\pi}^{\pi}\text{d}\theta (-1)^{l}J_{l}(\hat{\lambda}_{d0})J_{l'}(\hat{\lambda}_{d0}) e^{il(\theta-\theta_{0})}e^{il'(\theta-\theta_{0})}\\
    & = 2\pi\sum_{l'}(-1)^{l}J_{l}(\hat{\lambda}_{d0})J_{l'}(\hat{\lambda}_{d0})e^{-i(l+l')\theta_{0}}\delta(l_1+l')\\
    & = 2\pi\sum_{l}J_{l}^2(\hat{\lambda}_{d0})
\end{align*}
Then recall $ k_{\perp}\rho_L\ll 1 $ and $ k_{\perp}\rho_{d,\mathcal{E}}\ll 1 $, there is 
\begin{align}
    \left\langle e_i J_Z i\omega_d\delta H_0^{NL} \right\rangle_{Z} \propto & \hat{H}_0[J_1^2(\hat{\lambda}_0)-J_1^2(\hat{\lambda}_{0*})]\\
    = & k_{\theta,0}\hat{\rho}_{d,\mathcal{E}}^2(k_{r,0}+k_{r,0*})(k_{r,0}^2-k_{r,0*}^2)\notag\\
    = & k_{\theta,0}\hat{\rho}_{d,\mathcal{E}}^2 (k_{r,0}+k_{r,0*})^2(k_{r,0}-k_{r,0*})\notag\\ 
    = & -k_{\theta,0}k_Z^2\hat{\rho}_{d,\mathcal{E}}^2  i \hat{F}
\end{align}
where $ \hat{F}\equiv i(\hat{k}_{r,0}-\hat{k}_{r,0*})+\partial_r\ln \Phi_0-\partial_r\ln \Phi_{0*} $. Notice above result is a correction on Ref.\cite{qiu_effects_2016}. Putting it back into CCT, we have equation \eqref{E:CCTinZF}. 
\begin{align}
    &\left\langle e_i J_Z i\omega_d\delta H_0^{NL} \right\rangle_{\mathcal{E},Z} =\notag\\
    &\qquad -\dfrac{ce^2}{2B_0m}n_{0E}k_{\theta,0}k_Z^2\hat{\rho}_{d,f}^2 \hat{F}\hat{G}|\delta\phi_0|^2\label{E:CCTinZF}
\end{align}
where $ \hat{\rho}_{d,f}\equiv q v_f/\Omega_i$, $ \mathcal{E}_f=v_f^2/2 $, $ \hat{G} $ is defined as below
\begin{equation}
    \begin{split}
        \hat{G}\equiv & \bigg\langle \Omega_{*E}\dfrac{F_{0,E}}{2n_{0,E}} \dfrac{v_{\parallel,\mathcal{E}}^2}{v_f^2}\left(1-\dfrac{k_{\parallel}v_{\parallel}}{\omega}\right)_0\!\left(1-\dfrac{k_{\parallel}v_{\parallel}}{\omega}\right)_{0*}\\
        &\times\quad\sum_{l=\pm 1} \dfrac{-i}{\omega_0-k_{\parallel,0}v_{\parallel}-l\omega_{\rm{t}}-i\Omega_Z} \bigg\rangle_{\mathcal{E}}
    \end{split}
\end{equation}
with $ \Omega_{*E}\equiv\dfrac{cT_i}{eB_0}k_{\theta}L_{n_{0E}}^{-1} $. Meanwhile, the RS-MX contribution to ZM evolution is\cite{chen_nonlinear_2012,qiu_nonlinear_2017}:
\begin{equation}
    {\text{RS-MX}}=\dfrac{1}{2}\dfrac{e^2cn_0}{T_iB_0}k_{\theta,0}k_Z^2\rho_i^2\hat{F}\bigg(\!1-\dfrac{k_{\parallel,0}^2V_A^2}{\omega_0^2}\!\bigg)|\delta\phi_0|^2
\end{equation}

Then the evolution for ZF is:
\begin{align*}
        \left\langle (1-J_Z^2)F_{0,i} \right\rangle_{\mathcal{E}}&\partial_t\delta \phi_Z = \dfrac{cT_i n_{0E}}{2B_0 m} k_{\theta,0}k_Z^2\hat{\rho}_{d,f}^2\hat{G}\hat{F}|\delta\phi_0|^2 \\
        &+ \dfrac{1}{2}\dfrac{cn_0}{B_0}k_{\theta,0}k_Z^2\rho_i^2\hat{F}\left(1-\dfrac{k_{\parallel,0}^2V_A^2}{\omega_0^2}\right)|\delta\phi_0|^2
\end{align*}
which is written as equation (7) of the Letter:
\begin{equation*}
    \hat{\chi}_{iZ}\partial_t\delta \phi_Z= \dfrac{c}{B_0}k_{\theta,0}\hat{F}\!\left[\dfrac{n_{0E}}{n_0}\dfrac{\hat{\rho}_{d,f}^2}{\rho_i^2}\hat{G}+\!\left(1-\dfrac{k_{\parallel,0}^2V_A^2}{\omega_0^2}\right)\!\right]\!|\delta\phi_0|^2
\end{equation*}
where $ \left\langle (1-J_Z^2)F_{0,i} \right\rangle_{\mathcal{E}} = \frac{k_Z^2\rho_i^2n_0}{2}\hat{\chi}_{iZ} $ is used. Notice a factor $ 1-(k_{\parallel,0}^2V_A^2/\omega_0^2)\sim \varepsilon $ should arise from the integral of $ (1-\frac{k_{\parallel}v_{\parallel}}{\omega})_0(1-\frac{k_{\parallel}v_{\parallel}}{\omega})_{0*}$ in $ \hat{G} $ to characterize the breaking of ideal MHD condition. Then there is $ \hat{G}\propto \varepsilon f_r |\Omega_{*E}/\omega_0| $ with resonant particle fraction $ f_r $. The resulting ratio between CCT and RS-MX is 
	\begin{equation*}
		\dfrac{\rm CCT}{\text{RS-MX}}\sim \dfrac{f_rn_{0E}}{n_0}\dfrac{\hat{\rho}_{d,f}^2}{\rho_i^2}\left|\dfrac{\Omega_{*E}}{\omega_0}\right|\sim f_rq^2\dfrac{P_E}{P_i}\left|\dfrac{\Omega_{*E}}{\omega_0}\right| = q^2\dfrac{f_rn_{0E}}{n_0}\left|\dfrac{\hat{\omega}_{*E}}{\omega_0}\right|
	\end{equation*}
	where we defined $ \Omega_{*E}\equiv k_{\theta} L_{n_{0E}}^{-1}\frac{cT_i}{eB} $ and $ \hat{\omega}_{*E}\equiv k_{\theta} L_{n_{0E}}^{-1}\frac{cT_E}{eB} = \Omega_{*E}T_E/T_i $. Notice the last expression above is the same as $ \mathcal{O}(n_{E,R}\hat{\omega}_{*E}q^2/(n_0\omega_0\varepsilon)) $ obtained in Ref.\cite{qiu_nonlinear_2017} except the $ \varepsilon $ factor, which comes from the breaking of ideal MHD condition.
    \begin{itemize}
        \item For parameters state like in DIII-D at half minor radius: $ k_\theta=m/r\sim 5/0.25m $, $ 1/L_{n0E}\sim 1/0.25m$, $T_i=2keV, B=2T, q=1.5, f_r=0.25, f_{\beta}=1$, we have $ f_rq^2f_{\beta}|\Omega_{*E}/\omega_0|\sim 0.25\times1.5^2\times \frac{80kHz}{120kHz} =0.375<1$
        \item For parameters state like in JET at half minor radius: $ k_\theta=m/r\sim 5/0.45m $, $ 1/L_{n0E}\sim 1/0.45m$, $T_i=4keV, B=3.7T, q=1.5, f_r=0.25, f_{\beta}=1$, we have $ f_rq^2f_{\beta}|\Omega_{*E}/\omega_0|\sim 0.25\times1.5^2\times \frac{27kHz}{160kHz} \sim 0.1 <1$
    \end{itemize}
Thus, we conclude that
\begin{equation*}
    \dfrac{\rm CCT}{\text{RS-MX}}\ll 1
\end{equation*}

Another problem of keeping the CCT as a ZF drive source is we need to have a corresponding \textbf{\emph{sink}} in other physics quantities to satisfy the energy conservation. However, there is no explicit ``sink'' to supply energy to ZM in Ref. \cite{qiu_nonlinear_2017}. CCT accounts for the zonal contribution from curvature coupling with nonlinear EP fluctuation distribution. Therefore, the corresponding ``sink'' term is related to the EP distribution evolution, which might be constructed based on the complex theory in Refs.\cite{falessi_transport_2019,falessi_nonlinear_2023}. As we have assumed in the Letter, the EP profile is maintained by the sustained EP source. Considering that the contribution of the CCT is secondary and the model needs to conserve energy, we omit the CCT in equation (12) in the Letter to simplify the model. Meanwhile, because of the similar structures between CCT and RS-MX as we stated under equation (7) in the Letter, omitting CCT will not affect the validity of the simplified energy-conserving model.

Then the zonal mode evolution is dominated by RS-MX, and equation (7) of the Letter can be written as equation (12):
\begin{equation*}
    \hat{\chi}_{iZ}\partial_t\delta\phi_Z= \gamma_2|\delta\phi_0|^2-\nu\delta\phi_Z
\end{equation*}
with
\begin{equation}
    \gamma_2\equiv\dfrac{c}{B_0}k_{\theta,0}\hat{F}\left[\dfrac{n_{0E}}{n_0}\dfrac{\hat{\rho}_{d,f}^2}{\rho_i^2}\hat{G}+\left(1-\dfrac{k_{\parallel}^2V_A^2}{\omega_0^2}\right)\right]\simeq \dfrac{c}{B_0}k_{\theta,0}\hat{F}\varepsilon
\end{equation}
and $ \nu $ is the damping for zonal mode. The saturation level of TAE is
\begin{equation*}
    (|\delta\phi_0|^2)_S=\dfrac{\gamma_1\nu}{\gamma_d\gamma_2}\simeq \dfrac{\nu \dfrac{\pi\omega_A}{2\sigma}q^3k_{\theta,0}\rho_A\beta_Ef_r\dfrac{R_0}{L_{n0E}}}{2\dfrac{c^2}{B_0^2}k_{\theta,0}^2k_Z\hat{F}\varepsilon}
\end{equation*}
with $ \delta B_r=\frac{ck_{\theta}k_{\parallel}}{\omega_0}\delta\psi $ and $ \hat{F}\varepsilon=k_Z \hat{\chi}_{iZ}$, the saturation level of TAE fluctuating is 
\begin{align}
    (\dfrac{\delta B_r^2}{B_0^2})_S &=\dfrac{\pi q^2}{4\sigma\hat{\chi}_{iZ}}\dfrac{\nu\beta_Ef_r k_{\theta,0}}{\Omega_ik_Z^2L_{n0E}} \label{E:A:TAEsat-1}\\
    &\sim \dfrac{\pi q^2}{4\sigma\hat{\chi}_{iZ}\varepsilon^2}\dfrac{\beta_Ef_r k_{\theta,0}}{L_{n0E}}\dfrac{\nu\rho_i^2}{\Omega_i}
\end{align}
which is equation (21) of the Letter.

\section{EP profile evolution\label{A:EPevo}}
In deriving the evolution of the energetic particle distribution, we employ the quasi-linear method:
\begin{align*}
    \partial_t F_{0E}=&-\dfrac{c}{B_0}\Lambda_ZJ_{k'}\delta L_{k'}\delta H_{k''}^{L}\\
    =& \dfrac{ce}{B_0 m}\hat{H}_0J_{k_0}J_{k_{0*}}\delta L_{0}\delta L_{0*}Q_0F_{0E}\times \\
    & \qquad\bigg(\sum_{l*}\dfrac{J_{l*}^2(\hat{\lambda}_{d0*})}{\omega_{0*}-k_{\parallel,0*}v_{\parallel}-l_*\omega_{\rm{t}}-i\Omega_{Z*}}\\
    &\qquad\qquad+ \sum_{l}\dfrac{J_{l}^2(\hat{\lambda}_{d0})}{\omega_{0}-k_{\parallel,0}v_{\parallel}-l\omega_{\rm{t}}-i\Omega_{Z}}\bigg)\\
    =& -\dfrac{ce}{B_0 m}\hat{H}_0J_{k_0}J_{k_{0*}}\delta L_{0}\delta L_{0*}Q_0F_{0E}\times\\
    &\qquad\sum_{l=\pm 1}\dfrac{J_{l}^2(\hat{\lambda}_{d0})-J_{l}^2(\hat{\lambda}_{d0*})}{\omega_{0}-k_{\parallel,0}v_{\parallel}-l\omega_{\rm{t}}-i\Omega_{Z}}\\
    =&-\dfrac{ce}{B_0 m}k_{\theta,0}k_Z^2\hat{\rho}_{d,\mathcal{E}}^2\hat{F}J_{k_0}J_{k_{0*}}\delta L_{0}\delta L_{0*}Q_0F_{0E}\times\\
    &\qquad\sum_{l=\pm 1}\dfrac{-i}{\omega_{0}-k_{\parallel,0}v_{\parallel}-l\omega_{\rm{t}}-i\Omega_{Z}}\\
\end{align*}
Then after integral in $ \mathcal{E} $,
\begin{align}
    \partial_t n_{0E}&=-\dfrac{\pi}{2\sigma}\dfrac{c^2}{B_0^2}\dfrac{k_{\theta,0}^2k_Z^2\hat{F}\hat{\rho}_{d,f}^2|\delta\phi_0|^2f_r}{|\omega_0|}L_{n_{0E}}^{-1}n_{0E}\\
    &=-\dfrac{\pi}{2\sigma}\dfrac{k_Z^2\hat{\rho}_{d,f}^2f_r|\omega_0|}{k_{\parallel}^2}\bigg(\dfrac{\delta B_r}{B_0}\bigg)^2\hat{F}L_{n_{0E}}^{-1}n_{0E}\\
    &=\partial_r D_{\rm res}\partial_r n_{0E}
\end{align}
where
\begin{subequations}
    \begin{equation}\label{E:EPDres-Orig}
        \begin{split}
            D_{\rm res}=&\bigg\langle\dfrac{c^2}{B_0^2}k_{\theta,0}^2k_Z^2\hat{\rho}_{d,\mathcal{E}}^2\delta L_{0}\delta L_{0*}\dfrac{F_{0E}}{n_{0E}}\times\\
            &\sum_{l=\pm 1}\dfrac{-i}{\omega_{0}-k_{\parallel,0}v_{\parallel}-l\omega_{\rm{t}}-i\Omega_{Z}}\bigg\rangle_{\mathcal{E}}
        \end{split}
    \end{equation}
    \begin{equation}\label{E:EPDres}
        D_{\rm res}\simeq \dfrac{\pi}{2\sigma}\dfrac{k_Z^2\hat{\rho}_{d,f}^2V_Af_r}{k_{\parallel,0}}\bigg(\dfrac{\delta B_r}{B_0}\bigg)^2
    \end{equation}
\end{subequations}
is the equation (10) of the Letter.
Then, the evolution of EP profile is:
\begin{equation}\label{E:EPevo}
    \partial_t n_{0E}=-\gamma_3|\delta\phi_0|^2 n_{0E}+\text{Source} + \cdots
\end{equation}
with the external source and coefficient:
\begin{equation}
    \begin{split}
    \gamma_3 &=\dfrac{D_{\rm res}\hat{F}L_{n_{0E}}^{-1}}{|\delta\phi_0|^2}
    \end{split}
\end{equation}
Using equation (14) of the Letter, when the supply rate of external EP source exceeds the flatten rate of EP profile below, we can have a sustained EP profile.
\begin{equation}
    \dfrac{D_{\rm res}\hat{F}L_{n_{0E}}^{-1}}{\omega_A}= \dfrac{\pi}{4\sigma}f_r\dfrac{\hat{F}\hat{\rho}_{d,f}^2}{L_{n0E}}\hat{\nu}\hat{\gamma}_1
\end{equation}

\section{The evolution diagram of P-P system\label{A:dimensionlessPP}}
Dedimensionlize the equation (11) of the Letter:
\begin{align*}
    \partial_t|\delta\phi_0|^2&=\gamma_1 |\delta\phi_0|^2 - \gamma_d |\delta\phi_0|^2\delta\phi_Z\\
    \dfrac{1}{\omega_A}\partial_t\left|\dfrac{e\delta\phi_0}{T_i}\right|^2&=\dfrac{\gamma_1}{\omega_A} \left|\dfrac{e\delta\phi_0}{T_i}\right|^2 - \dfrac{\gamma_d}{\omega_A} \dfrac{T_i}{e} \left|\dfrac{e\delta\phi_0}{T_i}\right|^2\dfrac{e\delta\phi_Z}{T_i}
\end{align*}
and from equation (12) of the Letter there is:
\begin{align*}
    \hat{\chi}_{iZ}\partial_t\delta\phi_Z&=\gamma_2|\delta\phi_0|^2-\nu_{ii}\delta\phi_Z\\
    \dfrac{1}{\omega_A}\partial_t \dfrac{e\delta\phi_Z}{T_i}&=\dfrac{\gamma_2}{\hat{\chi}_{iZ}\omega_A} \dfrac{T_i}{e} \left|\dfrac{e\delta\phi_0}{T_i}\right|^2 - \dfrac{\nu_{ii}}{\hat{\chi}_{iZ}\omega_A}\dfrac{e\delta\phi_Z}{T_i}
\end{align*}
Then,
\begin{align}
    \partial_\tau x &= \hat{\gamma}_1 x - \hat{\gamma}_d x y \label{E:App-dlTAE-1}\\
    \partial_\tau y &= \hat{\gamma}_2 x - \hat{\nu}_{ii} y \label{E:App-dlZF-1}
\end{align}
where $ \tau\equiv \omega_A t $, $ x\equiv \left|\dfrac{e\delta\phi_0}{T_i}\right|^2 $, $ y\equiv \dfrac{e\delta\phi_Z}{T_i} $ and notice 
\begin{gather*}
    \Omega_{*E}=\dfrac{cT_i}{eB_0}k_{\theta}L_{n_{0E}}^{-1}=D_{B}k_{\theta}L_{n_{0E}}^{-1},\\
    D_B\equiv \dfrac{cT_i}{eB_0},\quad\hat{\chi}_{iZ}\sim 1.6\dfrac{q^2}{\sqrt{\varepsilon}} 
\end{gather*}
therefore,
\begin{align*}
    \hat{\gamma}_1&\equiv\dfrac{\gamma_1}{\omega_A} = \dfrac{\pi}{2\sigma}q^3k_{\theta,0}\rho_A\beta_Ef_r\dfrac{R_0}{L_{n0E}}\\ 
    \hat{\gamma}_d&\equiv\dfrac{\gamma_d}{\omega_A} \dfrac{T_i}{e}=\dfrac{2c}{B_0}k_{\theta,0}k_Z\dfrac{T_i}{e}\dfrac{1}{\omega_A}= \dfrac{2D_Bk_{\theta,0}k_Z}{\omega_A}\\
    \hat{\gamma}_2&\equiv\dfrac{\gamma_2}{\hat{\chi}_{iZ}\omega_A} \dfrac{T_i}{e}\simeq \dfrac{cT_i}{B_0e}\dfrac{k_{\theta,0}\hat{F}\varepsilon}{\hat{\chi}_{iZ}\omega_A} = \dfrac{D_Bk_{\theta,0}\hat{F}\varepsilon}{\hat{\chi}_{iZ}\omega_A}\simeq \dfrac{\hat{\gamma}_d}{2}\\
    \hat{\nu}_{ii}&\equiv\dfrac{\nu_{ii}}{\hat{\chi}_{iZ}\omega_A}
\end{align*}
The effective collisionless damping $ \nu_G $ can be incorporated into the expression $ \hat{\nu}=(\nu_{ii}+\nu_G)/ (\hat{\chi}_{iZ}\omega_A)$. Using equations \eqref{E:App-dlTAE-1} and \eqref{E:App-dlZF-1}, and decomposing the quantities into their saturated and fluctuating components $ x=x_0+\tilde{x} $, $y = y_0+\tilde{y} $, we can analyze the linear instabilities of this system.
\begin{align}
    \partial_\tau \tilde{x} &= \hat{\gamma}_1 \tilde{x} - \hat{\gamma}_d x_0 \tilde{y} - \hat{\gamma}_d \tilde{x} y_0 \\
    \partial_\tau \tilde{y} &= \hat{\gamma}_2 \tilde{x} - \hat{\nu}_{ii} \tilde{y}
\end{align}
Here $ x_0=y_0\hat{\nu}_{ii}/\hat{\gamma}_2 $, $ y_0=\hat{\gamma}_1/\hat{\gamma}_d $. The above system has two eigenvalues $ \lambda_{1,2} $:
\begin{equation}
    \lambda_{1,2}= -\dfrac{1}{2}\left[\hat{\nu}_{ii}\pm\sqrt{\hat{\nu}_{ii}(\hat{\nu}_{ii}-4\hat{\gamma}_1)}\right]
\end{equation} 
When $ \hat{\nu}_{ii}<4\hat{\gamma}_1 $, i.e. the damping of ZM is smaller than the growth of TAE $ \nu_{ii}\lesssim \gamma_{\rm TAE} $, the system will experience oscillatory decay to saturation states. The oscillation rate is determined from the imaginary part of eigenvalue $ |\Im \lambda_{1,2}|= \sqrt{\hat{\nu}_{ii}(4\hat{\gamma}_1-\hat{\nu}_{ii})}/2$, which is approximately $ f_{\rm osc}\sim \sqrt{\nu_{ii}\gamma_1}/\omega_A $.
\section{Geodesic acoustic transference\label{A:GAT}}
Here we introduce a simple generalization of geodesic acoustic transference \cite{scott_three-dimensional_1997,scott_geodesic_2003,scott_energetics_2005}. The evolution of ZF can be expressed as:
\begin{equation}
    \partial_t \langle \tilde{u}^{y}\rangle=\gamma_Z\langle \tilde{u}^{y}\rangle-\omega_B\langle p_s \sin \vartheta\rangle
\end{equation}
where $ \langle \tilde{u}^{y}\rangle $ is the ZF, $ \vartheta $ is the poloidal angle, $ p_s $ is the sideband pressure with $ (m=1,n=0) $, $ \gamma_Z $ represents the drive of ZF, $ \omega_B\equiv 2L_{\perp}/R $ is geodesic curvature coupling term from thermal plasma. The evolution of electron pressure sideband can be expressed as:
\begin{align*}
    \partial_t\langle p_{e,s} \sin \vartheta\rangle =&-\partial_x \left\langle \tilde{p}_e\tilde{v}_E^{x} \sin \vartheta\right\rangle + \langle v_{\parallel} \cos \vartheta\rangle \\
    &- \dfrac{\omega_B}{2} \left\langle \left(\partial_x p-\tilde{u}^{y}\right) \right\rangle\\
    \simeq & - \dfrac{D_{\rm turb}}{L_{\perp}^2}\langle p_{e,s} \sin \vartheta\rangle + \langle u_{\parallel} \cos \vartheta\rangle\\ 
    &-\langle J_{\parallel} \cos \vartheta\rangle-\dfrac{\omega_B}{2} \left\langle \partial_x p \right\rangle +\dfrac{\omega_B}{2} \left\langle \tilde{u}^{y}\right\rangle
\end{align*}
where the magnetic flutter effect is neglected, quasilinear turbulent coefficient is assumed, $ L_{\perp} $ is the perpendicular profile scale of thermals. The ion flow sideband $ \langle u_{\parallel}\cos \vartheta\rangle $ is typically small, current sideband and diamagnetic part cancel each other out. Then the evolution of pressure sideband is:
\begin{equation}
    \partial_t\langle p_{e,s} \sin \vartheta\rangle = -\dfrac{D_{\rm turb}}{L_{\perp}^2}\langle p_{e,s} \sin \vartheta\rangle +\dfrac{\omega_B}{2} \left\langle \tilde{u}^{y}\right\rangle
\end{equation}
Here we consider the simple case of damping through turbulent mixing and define $ \tau_{\rm turb}^{-1}\equiv (D_{\rm turb}/L_{\perp}^2) $, $ D_{\rm turb}\simeq \text{Re}\sum_{\tilde{\mathbf{k}}}i\frac{c^2}{B_0^2}|\phi_{\tilde{\mathbf{k}}}|^2/(\omega-\tilde{\mathbf{k}}_{\perp}\cdot\mathbf{v}_E) $, $ \tilde{\mathbf{k}}_{\perp} $ refers to thermal plasma turbulent wave number.

Dedimensionalizing the above equation, combining with the evolution system of ZF and TAE, we arrive at equations (17)-(19) of the Letter:
\begin{align}
    \partial_\tau x &= \hat{\gamma}_1 x - \hat{\gamma}_d x y \label{E:App-dlTAE-2}\\
    \partial_\tau y &= \hat{\gamma}_2 x -\hat{\nu}_{ii} y - \dfrac{C_S^2qR_0}{ik_ZD_B\hat{\chi}_{iZ}\omega_AL_{\perp}^2} \omega_Bz \label{E:App-dlZF-2}\\
    \partial_\tau z &= (1+\tau_i)\dfrac{ i k_ZD_B}{V_A}\dfrac{\omega_B}{2}y -\dfrac{1}{\omega_A \tau_{\rm turb}} z\label{E:App-dlPs-2}
\end{align}
where $ z\equiv \langle p_s \sin\vartheta\rangle=(1+\tau_i)\langle p_{e,s}\sin\vartheta\rangle $. It's easy to notice the geodesic acoustic oscillation:
\begin{equation}
   \dfrac{2(1+\tau_i)C_S^2/R^2}{\hat{\chi}_{iZ}\omega_A^2}=\dfrac{1}{\hat{\chi}_{iZ}}\dfrac{\omega_{\rm GAM}^2}{\omega_A^2}
\end{equation}
For the static sideband, there is
\begin{equation*}
    z_{\rm static} = (1+\tau_i)\dfrac{ i k_ZD_B}{V_A}\dfrac{\omega_B}{2}\omega_A\tau_{\rm turb}y 
\end{equation*}
Then the damping rate caused by geodesic acoustic transference is:
\begin{align}
    \hat{\nu}_{G}&=\dfrac{C_S^2qR_0}{ik_ZD_B\hat{\chi}_{iZ}\omega_AL_{\perp}^2} \omega_B z_{\rm static}\notag\\
    &=\dfrac{1}{\chi_{iZ}}\dfrac{\omega_{\rm GAM}^2}{\omega_A\tau_{\rm turb}^{-1}}
\end{align}
where $ \omega_{\rm GAM}^2\equiv 2(1+\tau_i)C_S^2/R^2=2q\omega_A^2\sqrt{\beta_{\rm bulk}} $. In the evolution of ZF, we can write down:
\begin{equation}
    \partial_{\tau} y=\hat{\gamma}_2 x -\hat{\nu}_{ii} y -\hat{\nu}_{G} y
\end{equation}
With $ \omega_A\sim 10^{6} \,\rm{rad/s}$, $ \omega_{\rm GAM}\sim 10^{5}\, \rm{rad/s}$, and $ \tau_{\rm turb}^{-1}=D_{\rm turb}/L_{\perp}^2\sim 10^{5} \,\rm{Hz}$, we have:
\begin{equation}
    \hat{\nu}_{G} \sim \mathcal{O}(10^{-2})-\mathcal{O}(10^{-1})
\end{equation}

\section{Discussion on phase space relaxation}
The phase-space gradient relaxation is indeed important for the AE saturation, which we neglected in this Letter. However, for the problem of cross-transport including EP, AE, ZF and turbulence, the phase space transport can be mitigated. One signature of strong phase-space activities is the mode chirping and overlapping in spectrum. We notice that the phenomena of ``EP-induced ITB'' is usually accompanied by ``single AE'' spectrum, where spectrum of modes usually consist of several separated AEs and there is no strong chirping and overlapping\cite{mazzi_enhanced_2022,garcia_stable_2024-1}. The reason behind this could be the scattering and mixing of clumps and holes by the existence of turbulence \cite{liu_cross-scale_2024}. We know from  Refs.\cite{todo_nonlinear_2010,chen_zonal_2018,biancalani_effect_2020,biancalani_gyrokinetic_2021} that when AE generate ZM, the saturation level of AE will be reduced. Based on Ref. \cite{berk_scenarios_1992}, the effective ``rate of reconstruction of the unperturbed distribution function after it has been flattened in phase space by a nonlinear wave'' is $ \nu_{\rm eff}\approx \nu_{\perp} \omega/\omega_b $. Since the trapping frequency of particle in a wave is $ \omega_b\propto \phi_{\rm AE}^{1/2} $, with a reduced AE amplitude, there will be a higher restoration rate in phase space for EP. Therefore, the existence of ZM can also mitigate the phase space clumps and holes.

The existing phase-space zonal structure theories remain complex and has yet to provide any clear predictions or quantitative/semi-quantitative estimates regarding EP-induced ITBs as we did in this work. The theory presented in this Letter can also serve as a foundation for more accurate models that include phase-space dynamics in the future.

\section*{The parameters used to generate Figure 2 and Figure 3}
We used a combination of DIII-D parameters and profiles similar to those in Figure 1 of Ref. \cite{todo_nonlinear_2010}, as listed in Table \ref{T:1}. The derived parameters used to generate Figures 2, 3, and other estimates are provided in Table \ref{T:2}. Note that the  $ 2\pi $  factor is neglected in all frequency calculations.

\begin{table}[htbp]
    \centering\begin{tabular}{l|c|r}
        $ R_0 $        & 1.75                          & m\\
        $ a $          & 0.64                          & m\\
        $ \varepsilon $& $a/R_0$                       & 1\\
        $ L_{n0E} $    & 0.5$a$                          & m\\
        $ L_{\perp} $  & 0.9$a$                          & m\\
        $ B_0 $        & 20                            & kGs\\
        $ n_0 $        & $3\times 10^{13}$             & cm$^{-3}$\\
        $ T_{\rm ev} $ & $2\times 10^3$                & eV\\
        $ f_{\beta} $  & 1                             & 1\\
        $ \tau $       & 1                             & 1\\
        $ n $          & 4                             & 1\\
        $ m $          & 6                             & 1\\
        $ q $          & $(m+1/2)/n$                   & 1\\
        $ c $          & $3\times 10^{10}$              & cm/s\\
        $ e $          & $4.8\times 10^{-10}$            & esu\\
        $ V_f/V_A $    & 1.2                            & 1\\
    \end{tabular}
    \caption{The parameters are based on the DIII-D device and the EP profile in Ref. \cite{todo_nonlinear_2010}.}
    \label{T:1}
\end{table}

\begin{table}[htbp]
    \centering\begin{tabular}{l|c|r}
        $\rho_i$        & 0.0031305                     & m\\
        $\Omega _i$     & $3.04845\times 10^7$          & 1/s\\
        $k_{\perp}=0.5 \epsilon/\rho _i$     & 58.4116  & 1/m\\
        $k_{\theta 0}$  & 10.1563                       & 1/m\\
        $k_Z=k_{\perp}/4$& 14.6029                      & 1/m\\
        $f_{\beta}$     & 1                             & 1\\
        $f_{r}$         & 0.25                          & 1\\
        $\beta_{\rm bulk}$& 0.0120697                   & 1\\
        $\sigma =\left(\frac{V_f}{V_A}\right)^2 \log \left(\left(\frac{V_f}{0.5 V_A}\right){}^2\right)$        & 2.5                           & 1\\
        $D_B$           & 1000.                         & m$^2$/s\\
        $V_A$           & $8.03326\times 10^6$          & m/s\\
        $C_s$           & $200000 \sqrt{5}$             & m/s\\
        $\omega_A$      & $2.82488\times 10^6$          & 1/s\\
        $\omega_{\rm GAM}$& 511101.                     & 1/s\\
        $\Omega_{*E}/\omega_A$& 0.0112353                & 1\\
        $\nu_{ii}$      & 391.284                       & 1/s\\
        $\nu_{\rm total}$& $2.04839\times 10^6$         & 1/s\\
        $\omega_{\rm ExB}$& 210204.                      & 1/s\\
        $\omega_{\rm ExB}a/C_s$& 0.300819                & 1\\
        $\Omega_Z=k_{\theta 0}V_{\rm ExB}$& 18274.5      & 1/s\\
        $(\delta B_r/B_0)^2$& $2.22057578350852\times 10^{-7}$& 1\\
        $\text{TAE}_{\rm sat/th}$& 0.268055                    & 1\\
        $\hat{\gamma}_1$& 0.118065                     & 1\\
        $\hat{\gamma}_d$& 0.105003                     & 1\\
        $\hat{\gamma}_2$& 0.0525015                    & 1\\
        $\hat{\nu}_{ii}$& 0.000019826                  & 1\\
        $\hat{\nu}_G$   & 0.10377                      & 1\\
        $\hat{\gamma}_{\rm G1}$& $-3.91558 i$           & 1\\
        $\hat{\gamma}_{\rm G2}$& $+0.00119663 i$        & 1\\
        $\hat{\gamma}_3$& 0.0451527                    & 1\\
        $1/\tau_{\rm turb}$& 127551.                     & 1/s\\
        $1/(\omega_A\tau_{\rm turb})$& 0.0451527       & 1\\
        $3\hat{\gamma}_d\hat{\gamma}_2-(\omega_A\tau_{\rm turb})^{-2}$& 0.0144997& 1\\
        $\tau_Z/\tau_{\rm turb}$& 0.860026               & 1\\
        $f_{\rm osc}$   & 0.110698                     & 1/$\omega_A$\\
        $(f_rf_{\beta})_{\rm crit}$& 0.0831064           & 1
    \end{tabular}
    \caption{A series of parameters derived from Table 1, used for Figures 2, 3, and estimates in the Letter. Note that the  $ 2\pi $ factor is neglected in all frequency calculations.}
    \label{T:2}
\end{table}

\newpage
\bibliographystyle{unsrt}
\bibliography{refs}
